\documentclass[journal,hideappendix]{vgtc}        

\onlineid{0}

\vgtccategory{Research}

\vgtcpapertype{Application/Design Study}

\title{AgentCoord: Visually Exploring Coordination Strategy for LLM-based Multi-Agent Collaboration}

\author{%
  Bo Pan, Jiaying Lu, Ke Wang, Li Zheng, Zhen Wen, Yingchaojie Feng, Minfeng Zhu, and Wei Chen
}

\authorfooter{
  \item
    Bo Pan, Jiaying Lu, Ke Wang, Li Zheng, Zhen Wen, Yingchaojie Feng, and Wei Chen are with the State Key Lab of CAD\&CG, Zhejiang University, and Wei Chen is also with the Laboratory of Art and Archaeology Image (Zhejiang University), Ministry of Education, China.
   E-mail: \{bopan\,$|$\,ljying\,$|$\,sttot$|$\,zju\_zhengli\,$|$\,wenzhen\,$|$\,fycj\,$|$\,chenvis\}@zju.edu.cn.

    \item
      Minfeng Zhu is with Zhejiang University.
      E-mail: minfeng\_zhu@zju.edu.cn.

}

\abstract{
The potential of automatic task-solving through Large Language Model (LLM)-based multi-agent collaboration has recently garnered widespread attention from both the research community and industry. While utilizing natural language to coordinate multiple agents presents a promising avenue for democratizing agent technology for general users, designing coordination strategies remains challenging with existing coordination frameworks. This difficulty stems from the inherent ambiguity of natural language for specifying the collaboration process and the significant cognitive effort required to extract crucial information (e.g. agent relationship, task dependency, result correspondence) from a vast amount of text-form content during exploration. In this work, we present a visual exploration framework to facilitate the design of coordination strategies in multi-agent collaboration. We first establish a structured representation for LLM-based multi-agent coordination strategy to regularize the ambiguity of natural language. Based on this structure, we devise a three-stage generation method that leverages LLMs to convert a user’s general goal into an executable initial coordination strategy. Users can further intervene at any stage of the generation process, utilizing LLMs and a set of interactions to explore alternative strategies. Whenever a satisfactory strategy is identified, users can commence the collaboration and examine the visually enhanced execution result. We develop AgentCoord, a prototype interactive system, and conduct a formal user study to demonstrate the feasibility and effectiveness of our approach.

}


\teaser{
  \centering
  \includegraphics[width=\linewidth, alt={A view of a city with buildings peeking out of the clouds.}]{figs/teaser.pdf}
  \caption{%
  Our visual exploration framework for coordination strategy design: The user first enters a general goal for agent collaboration (a). The system conducts a three-stage generation (i.e., Plan Outline Generation (b), Agent Assignment (c), and Task Process Generation (d)) to provide an initial strategy. The user interactively explores alternative strategies for each stage with the help of LLMs (e, f, g). Once satisfied, the user commences the collaboration and examines the visually enhanced execution results (h).
  }
  \label{fig:teaser}
}

\graphicspath{{figs/}{figures/}{pictures/}{images/}{./}} 

\usepackage{tabu}                      
\usepackage{booktabs}                  
\usepackage{lipsum}                    
\usepackage{mwe}                       

\usepackage{amsmath}
\usepackage{amssymb}
\usepackage{mathtools}
\usepackage{graphicx}
\begin{document}


\maketitle

\definecolor{bdcolor}{RGB}{68,170,168} 
\newcommand{\overviewBox}[1]{%
\tikz[baseline=-0.5ex]{\node[rounded corners=0.2em, 
fill=white, 
draw=bdcolor, 
minimum size=0.9em, 
text=bdcolor, 
text centered,
inner sep=0pt,
line width=0.9pt,
font=\fontsize{9pt}{0.9em}\fontseries{ul}\selectfont](TsNode){#1}}}

\section{Introduction}
Large Language Model (LLM) based agents, which are capable of observing, making decisions, and performing actions with the reasoning capabilities of LLM, have undergone significant improvements, showing great potential in various areas such as programming\cite{hong2023metagpt,chetDev}, creative writing\cite{AutoAgents, wang2023unleashing, OKRagents}, and question answering \cite{wu2023empiricalMath, tang2023MedAgents}. 
While initial forays into the realm of LLM-based agents focused on solitary agent system\cite{AutoGPT,Expertprompting}, the concept of multi-agent collaboration—mirroring the cooperative interactions among humans—has started to pique the interest of the AI research community. Drawing inspiration from the synergistic outcomes observed in human teamwork\cite{engelbart2023augmenting, woolley2010evidence,luppi2022synergistic}, a burgeoning body of research has been investigating and validating the benefits (e.g. expand expertise\cite{Salewski2023InContextIR, tang2023MedAgents}, enhance reliability \cite{chan2023chateval, du2023improving_debate, chetDev}, encourage divergent thinking\cite{Encouraging_Divergent_Thinking_Debate, zhuge2023mindstorms}) brought about by LLM-based multi-agent collaboration.

In order to facilitate the coordination of multi-agent collaboration, the open-source community emerges a multitude of frameworks for prototyping LLM-based multi-agent systems. Existing multi-agent frameworks can be divided into two categories based on how users can specify or intervene in the collaborative process: code-based and natural language-based. For code-based frameworks \cite{hong2023metagpt, chetDev, wu2023autogen, CrewAI, Langroid, chen2023agentverse}, users need to hard-code the coordination strategies (e.g. the division of tasks, the assignment of agents, the flow of massages) directly into the code, which requires code skill and learning cost. The natural language-based frameworks\cite{AutoAgents, wu2023autogen} \footnote[1]{{AutoGen \cite{wu2023autogen} supports both code-based and natural language-based paradiam. In its ``group chat mode'', the coordination strategy can be expressed in free-form natural language and coordinated by a chat manager.}}, which directly uses natural languages to specify the coordination strategies, could be a promising way to democratize agent technology for broader general users. Additionally, given that LLMs inherently possess coordinating capabilities and rich domain knowledge across different tasks, the natural language-based frameworks can easily leverage LLMs to assist in drafting and refining the coordination strategies represented in natural language\cite{AutoAgents}. 

However, designing coordination strategies remains challenging with existing natural language-based frameworks. First, the flexibility of natural language could be a double-edged sword: on one hand, it allows users to freely design and express their coordination strategies; on the other hand, overly flexible expressions can make the devised collaboration strategies ambiguous, often requiring users repeatedly engage in remedial specification to ensure the execution of the collaboration doesn't stray from its intended course. Second, representing and exploring coordination strategies in pure text format faces challenges as the complexity of the collaboration process and team organization rises. Users can easily get lost in the ``text jam'' where important information (e.g. agent relationship, task dependency, result correspondence, strategy discrepancy) they care about at certain points during exploration is drowned in blocks of text. Therefore, novel approaches are highly desirable to enhance the current natural language-based design process.

This work thus presents a visual exploration framework for efficiently designing coordination strategies for LLM-based multi-agent collaboration. To exploit the flexibility of natural language while incorporating a level of organization to regularize its ambiguity, we analyze the common concepts and structures found in the description for coordination strategy in a corpus of 25 LLM-based multi-agent collaboration papers and 7 high-star projects \footnote[2]{{The corpus can be found in our project repository.}}, based on which we establish a structured representation for LLM-based Multi-agent coordination strategy. This structure lays a foundational scaffolding for the entire exploration process. Based on this structure, we design a generation method that exploits the coordinating capabilities and domain knowledge of LLMs to map the general goal provided by users (\cref{fig:teaser} \overviewBox{a}) into an executable initial coordination strategy to help users kick off the exploration. To ensure coherence among various parts of the generated strategy, we divide the generation process into three stages: Plan Outline Generation (\cref{fig:teaser} \overviewBox{b}) for an overall collaboration plan to achieve the goal, Agent Assignment (\cref{fig:teaser} \overviewBox{c}) for each task in the plan outline, and Task Process Generation (\cref{fig:teaser} \overviewBox{d}) for specifying how assigned agents collaboratively finish the task. To streamline users' exploration and iterative refinement for alternative strategies, we propose a set of interactions (\cref{fig:teaser} \overviewBox{e} \overviewBox{f} \overviewBox{g}) to help users visually explore the design space for each generation stage with the help of LLMs. Whenever users are satisfied with a certain strategy, they can initiate the collaboration and examine the execution result which is visually enhanced and linked with previous stages for efficient verification (\cref{fig:teaser} \textbf{\overviewBox{h}}). 

To validate the feasibility and effectiveness of this framework, we developed an interactive system called \textit{AgentCoord} that enables users to visually explore coordination strategies for LLM-based multi-agent collaboration, effectively integrating the prior of LLMs and users during design. Our user study, involving 12 users with a general interest in LLM-based multi-agent collaboration, suggests that our approach can effectively facilitate the design process for LLM-based multi-agent coordination strategies and has the potential to democratize agent coordination for broader users. In summary, our contributions include:

\begin{itemize}
\item A visual exploration framework that enables general users to efficiently design
coordination strategy for LLM-based multi-agent collaboration.
\item \textit{AgentCoord} \footnote[3]{{Project Repository: https://github.com/AgentCoord/AgentCoord}}, an open-source interactive system that instantiates our framework with a set of interactions and visual designs to facilitate coordination strategy exploration.
\item A formal user study that
demonstrates the feasibility and effectiveness of our approach.
\end{itemize}

\section{Related Work}
\subsection{LLM-Based Multi-Agent Collaboration}
Large language models (LLMs) have recently demonstrated impressive capabilities as versatile task-solving agents, attracting substantial interest in both industry and academia \cite{wang2023survey,xi2023rise}. Since LLMs are trained on natural language corpus that is biased toward human thinking \cite{zhuge2023mindstorms} and optimized for conversation \cite{ouyang2022training}, LLM-based agents can collaborate through natural language in a human-like manner and harness a range of benefits that come with collaboration.

Recent works have initiated attempts to coordinate agents with varied expertise in order to improve outcomes on a wide spectrum of tasks that benefit from a diversity of knowledge. Medagent \cite{tang2023MedAgents} collects medical agents with different specialties to provide a comprehensive analysis of patient's conditions and treatment options. MetaGPT \cite{hong2023metagpt} and ChatDev \cite{chetDev} enable agents with different roles such as product managers, designers, and programmers to collaborate in software development, thereby improving the quality of the software produced. MARG\cite{darcy2024marg} develops a framework for integrating the proficiency of multiple expert agents to review scientific papers. AutoAgents\cite{AutoAgents} and OKR-Agent\cite{OKRagents} demonstrate how creative content tasks, such as creative writing and storyboard generation, can benefit from the collaboration of agents with diverse domain backgrounds. 
AgentVerse\cite{chen2023agentverse} showcases a scenario in which multiple experts with different backgrounds collaborate to deliver a hydrogen storage station siting solution. 

Additionally, recent studies have discovered that multiple agents can work together to foster cognitive synergy \cite{luppi2022synergistic} similar to humans. Chan et al. request multiple agents to delve into the discussion from diverse perspectives to catalyze a comprehensive assessment that is greater than the sum of their separate assessments. Liang et al.\cite{Encouraging_Divergent_Thinking_Debate} and Du et al.\cite{du2023improving_debate} let multiple agents to debate with each other to encourage deeper levels of contemplation. Zhuge et al.\cite{zhuge2023mindstorms} propose the concept of ``mindstorm'' to describe how multiple agents take multiple rounds of communication to iterate the ideas to find a solution that is often superior to any individual solution. 

Despite the potential of multi-agent collaboration shown in various fields, most works require coding to design coordination strategies for agents, limiting accessibility for broader general users. While AutoGen \cite{wu2023autogen} and AutoAgents \cite{AutoAgents} support representing coordination strategies in pure natural language, users still encounter a series of issues when using natural language to explore and design coordination strategies. Our work attempts to address these issues with structured generation of coordination strategies and a set of visualization approaches to facilitate users' understanding and exploration of the coordination strategies.

\subsection{Generating Coordination Strategy using LLMs}
Although LLM-based agents have shown tremendous potential in collaboratively completing tasks, manually designing coordination strategies is often challenging, time-consuming, and sometimes requires expertise in specific domain knowledge \cite{li2023camel}. Thus, it is highly desirable to leverage LLM's inherent coordinating capabilities and prior knowledge across different tasks to aid in designing coordination strategies for agent collaboration.

Many works on multi-agent collaboration leverage the prior knowledge of LLMs for the formation and adjustment of agent teams. Wang et al. \cite{wang2023unleashing} prompt LLM to dynamically identify a set of agent roles in response to a task query. Medagent \cite{tang2023MedAgents} prompts LLM to work as a medical expert who specializes in categorizing a specific medical scenario into specific
areas of medicine and gathering corresponding expert agents. DyLAN \cite{liu2023dynamic} leverages LLM to score the performance of agents and dynamically optimize the team organization during collaboration. The AgentBuilder module of Autogen \cite{wu2023autogen} prompts LLM to generate system prompts for multiple agents based on the current task and add them to a group chat for collaboration. AutoAgents \cite{AutoAgents} designs an LLM-based Agent Observer to check the compliance of the agent with the requirements and make suggestions for adjustments.

LLMs are also widely utilized to aid in planning the collaboration process of multiple agents. For example, AutoAgents \cite{AutoAgents} prompt LLMs to draft a collaboration plan that specifies the agents involved in each step and the expected outputs. In Autogen's group chat mode \cite{wu2023autogen}, the user can play the role of an Admin agent to draft and refine the collaboration plan together with an LLM-based Planner agent. OKR-agent \cite{OKRagents} leverage LLMs to recursively decompose the tasks for teams of agents. Further, AgentVerse \cite{chen2023agentverse} designs a collaborative decision-making stage for multiple LLM-based agents to make short-term planning.

Our work focuses more on how to leverage LLM to facilitate general users in designing their own multi-agent coordination strategy. To achieve this, we propose an LLM-based three-stage generation method to generate a structured coordination strategy based on the user's goal. Additionally, we propose a set of interactions to assist users in flexibly exploiting the coordination ability of LLMs during exploration.

\subsection{Interface for LLM-based Agents}
During the execution of LLM-based agents, a multitude of intricate information is involved, which is hard to digest with a plain text terminal \cite{weng2024insightlens,lu2024agentlens}. Therefore, it is highly desirable to have some interfaces to assist in understanding and intervening in the execution process. Early interfaces \cite{AgentGPT, xagent2023, SuperAGI} for monitoring single LLM-based agents typically feature an outline view that maps out the overall execution process, complemented by detailed text blocks enhanced with highlighting and icons. For systems\cite{park2023generative, lin2023agentsims, CooperativeEmbodiedAgent, chetDev} that deploys multiple agents in a virtual sandbox environment, a panoramic view is usually provided to transform text-form information into concrete visual elements (e.g. moving agent avatars, expressive emojis) for easier comprehension of the overall process. Topological structures such as trees and graphs are also utilized to visualize and manage the execution processes of agents. SPROUT\cite{liu2023sprout} employs a tree structure to assist users in visualizing and controlling the process of an agent composing code tutorials. Hong et al. \cite{hong2024data} use a hierarchical graph structure to manage the execution process of a data science agent and allow users to interactively edit the graph during execution. AutoGen \cite{wu2023autogen} introduces a transition graph to allow users to constrain agent transition to mitigate the risk of sub-optimal agent transitions during multi-agent collaboration in Group Chat mode. Recently, AgentLens \cite{lu2024agentlens} initiated the first attempt to design a visual analysis system to assist users in analyzing the agent behaviors in LLM-based multi-agent systems.

Extending this line of work, our work enables general users to visually explore coordination strategies for LLM-based multi-agent collaboration.

\section{Formative Study}
To gain insight into how users use current natural language-based frameworks to coordinate multiple LLM-based agents and identify the challenges that exist during the exploration process for coordination strategy, we carried out a formative study. Based on the findings in the formative study, we formulated four design requirements to enhance the process of designing coordination strategies for LLM-based multi-agent collaboration.
\subsection{Participants and Procedure}

We recruited 8 participants who have experience or general interest in LLM-based multi-agent collaboration from the local university and online discussion platforms for open-source multi-agent frameworks. Four of them are experienced experts in LLM-based multi-agent systems (E1 and E2 are NLP researchers familiar with LLM-based multi-agent collaboration, while E3 and E4 are developers having experience in constructing multi-agent systems). Another four of them (G1-4) are general users who have a basic understanding of LLM-based agents and are interested in building their own LLM-based multi-agent collaboration strategy. 

\textbf{Procedure:} In our formative interviews,
we initially asked the participants about their prior experience with any LLM-based multi-agent framework or system. Afterward, we show participants how to use natural language to specify the coordination strategy for multi-agent collaboration using AutoAgents \cite{AutoAgents} alongside a text editor and the ``group chat mode'' of AutoGen\cite{wu2023autogen}. After the participants got familiar with the usage, they were asked to choose a task that could benefit from collaboration among multiple LLM-based agents and construct their own coordination strategy with both systems separately. The participants can also use ChatGPT to assist them in designing coordination strategies during the process. Finally, we gathered feedback on participants' experience during the construction of the coordination strategy and inquiry about the challenges they faced during the process. For participants who reported prior experience with any LLM-based multi-agent framework at the start of the interview, we also asked them to compare the strengths and weaknesses of natural language-based methods with previous frameworks they have used.

\subsection{Findings}
All participants find using natural language to design coordination strategies is intuitive and could be a promising approach to democratizing agent coordination for a wider general audience. However, several challenges are also identified, which hinder the participants' current exploration process to design coordination strategies at their will.

\textbf{Lack of structure to regularize the ambiguity of natural language.} 
Although natural language is easy to understand and highly expressive, it is prone to ambiguity. For example, for a high-level cooperation strategy description ``\textit{Pharmaceutical Chemist, Patent Agent, and Clinical Research Scientist collaboratively drafting a patent application}'', there could be many ambiguous aspects: ``\textit{Who takes the main responsibility for drafting?}'', ``\textit{How are opinions integrated?}''... During the formative study, we notice that users often start with a generated collaborative strategy and then, upon observing the unexpected outcomes, identify areas that were not articulated and make remedial enhancements to the original cooperation strategy. After several rounds, the collaborative strategy specification ``\textit{could be lengthy and messy to read}'' (G4), and ``\textit{sometimes even contain self-conflicts}'' (G3). Both E1 and E4 suggest that ``\textit{some structures should be provided to regularize the design process}'' (E1, E4), which can ``\textit{draw on designs from current code-based frameworks}'' (E1).

\textbf{Lost in the vast amount of intricate text.} During the process of designing collaborative strategies, users need to refer to a substantial amount of text information (e.g., previously designed collaborative strategies, descriptions of different agents, input/output of agents, intermediate objects). The vast amount of intricate text poses significant cognitive overhead on users during the design process. Many participants express the feeling that the quantity of text is ``\textit{overwhelming}'' (E3, E4, G1-G4). They usually needed to ``manually switch back and forth between different parts of the texts'' (G2) and ``sometimes forget where to find'' (G3). E3 mentioned that ``\textit{as the complexity of my strategy rises, maintain a clear connection between specific execution result and the corresponding part of my strategy become challenging}'' (E3). The Participants also express the need to ``\textit{have a visual interface that helps organize information}'' (G2, E3).

\textbf{Lack of interactions support to facilitate exploration.} To leverage the powerful coordination capabilities of LLMs and deal with ``writer’s block'', participants often chat with LLMs (e.g. using ChatGPT) to aid them in drafting and exploring coordination strategies. However, the linear non-reversible conversation interface for chatting is not designed for iterative multi-thread exploration. E2 mentioned that  ``\textit{managing exploration history and toggling between different possibilities with manual copy \& paste is cumbersome}'' (E2). Additionally, participants also mentioned concerns over the fluidity of exploration due to the need for ``\textit{manually craft auxiliary prompts for different exploration purposes}'' (G2). Moreover, due to the stochastic nature of LLM outputs (which is also greatly affected by prompts) and diverse possibilities for strategy design, participants express the need for ``\textit{an interface to help systematically explore and compare different outputs by LLM}'' (G1). 

\subsection{Design Requirement}
In response to the problems identified in the formative study, our goal is to develop an interactive system to help general users smoothly explore and design coordination strategies for LLM-based Multi-agent collaboration. The design requirements are summarized as follows:

\textbf{R1: Generate a structured coordination strategy for the user's goal.} Although using natural language to describe a coordination strategy lowers the barrier for users and brings a high level of flexibility, users often lack an effective way to deal with its ambiguity. Therefore, the system should provide a structure of coordination strategy to help regularize the ambiguity inherent in natural language and serve as a scaffolding for downstream exploration. Moreover, to help the user kick off, the system should be able to generate an initial coordination strategy based on the user's goal leveraging the coordination capability of LLMs.

\textbf{R2: Provide an effective visual organization for the strategy.} When users are devising a coordination strategy, they need to refer to various types of relevant information represented in the text. However, at present, users have to flip back and forth through a vast amount of plain text to search for target information and do verification, which creates a significant cognitive load. Therefore, the system should effectively visually organize and enhance the various pieces of information involved in the coordination strategy design process to help users quickly locate the information they need and provide the relevant context.

\textbf{R3: Support flexible interactions to facilitate strategy exploration.} Users often need to explore various possible options at different stages of the coordination strategy design with the help of LLMs; however, iterative exploration based on a linear chat interface with LLMs is cumbersome and unintuitive. Therefore, the system should support flexible and intuitive interactions to help users conduct multi-thread iterative exploration. Moreover, the system needs to offer assistance to help systematically explore and compare different design choices for coordination strategies.

\textbf{R4: Provide visual enhancement for the execution result.} During the execution of collaborative tasks, agents generate a substantial amount of textual information. However, at present, both code-based and natural language-based agent coordination frameworks only output results through a plain text terminal. Users often have to manually switch back and forth between different parts of the coordination strategy and the execution results to establish connections, which increases cognitive load and decreases analytical efficiency. Therefore, the system needs to provide visual enhancements to help users examine execution results.

\section{Structured Coordination Strategy Generation}
\label{section 4}
In this section, we abstract a structured representation for coordination strategy design (Section \ref{sec4.1:Common Structure for Coordination Strategy}) to regularize the ambiguity of natural language \textbf{(R1)}. Based on this structure, a three-stage generation method (Section \ref{sec4.2:Three-stage Strategy Generation}) has been designed to automatically generate an initial coordination strategy based on the user's goal \textbf{(R1)}.
\subsection{Structured Representation for Coordination Strategy}
\label{sec4.1:Common Structure for Coordination Strategy}
To maximize the expressiveness of the structure defined for coordination strategy, E1-4 and we collaboratively survey a corpus of 25 LLM-based Multi-agent collaboration papers and 7 high star open source frameworks \cite{hong2023metagpt, li2023camel,wu2023autogen,CrewAI,AutoAgents,chetDev,chen2023agentverse} for Multi-agent coordination. We analyze the common concepts and structures found in the description for coordination strategy in those papers and projects, based on which we establish a common structure for LLM-based Multi-agent coordination strategy. 

We depict the relationship between key concepts involved in the coordination strategy structure as follows:

\begin{itemize}

\item \textbf{Plan Outline:} Provides a blueprint for the overall collaboration, typically breaking the objectives down into a sequence of \textbf{Tasks} to be carried out one after the other.

\item \textbf{Task}: Takes \textbf{Key Objects} as input and output its target \textbf{Key Object}. The task process specifies how \textbf{Agents} collaboratively finished the task.

\item \textbf{Key Object}: Important intermediate objects during collaboration. 

\item \textbf{Agent}: Intelligent entity that can perform \textbf{Action} according to its observation and \textbf{Instruction}. 

\item \textbf{Action}: The smallest unit of agent behavior observable.
 
\item \textbf{Instruction}: Natural language-based specification that tells agent what/how to do.

\end{itemize}

\subsection{Three-stage Strategy Generation}
\label{sec4.2:Three-stage Strategy Generation}
To help users kick off the exploration, we design a three-stage generation method to provide an initial coordination strategy based on the goal of the user, leveraging the coordination capability of LLM. Details of all the prompts used can be found in our project repository.

\subsubsection{Stage1: Plan Outline Generation}
Given a general description of the goal $g$ provided by the user and a set of initial key objects $\mathcal{I} = \{ko_{1},ko_{2}, ..., ko_{n}\}$, the goal of the Plan Outline Generation stage is to draft a plan outline that decomposes the final goal into a sequence of step tasks $\mathcal{P} = \{t_{1},t_{2}, ..., t_{n}\} = \{t_{i}\}|_{i=0}^{n} $. Here each task $t_{i}$ contains the following attributes:
\begin{itemize}

\item \textbf{Step Name:} A clear and concise name summarizing the step.
\item \textbf{Task Content:} Task description of the current step.
\item \textbf{Input Object List:} The input key objects that will be used in the current step.
\item \textbf{Output Object:} The output key object of the current step.

\end{itemize}

To achieve this, we prompt the LLM to serve as an expert plan outline designer to carefully analyze and decompose the goal provided by the user and output the plan outline $\mathcal{P}$:

\begin{equation}
         \mathcal{P} = \text{LLM} ( g, \mathcal{I}, \texttt{prompt}_\texttt{stage1}) 
\end{equation}

\subsubsection{Stage2: Agent Assignment}
After the plan outline is generated, a team of agents $\mathcal{A}_{i} = \{agent_{1}, agent_{2}, ..., agent_{n\}}$ should be assigned for each task $t_{i}$. Those agents are selected from a board of agent candidates $\mathcal{AB} = \{agent_{1}, agent_{2}, ... agent_{m}\} (\mathcal{A}_{i} \subseteq \mathcal{AB} )$ provided by user. The agents in agent board $\mathcal{AB}$ can be obtained through role prompting\cite{Expertprompting}, LLM fine-tuning\cite{wei2021finetuned}, retrieval-augmented generation (RAG) \cite{lewis2020retrieval}, or even recruitment from an agent store\cite{NexusGPT,GPT-Store,SuperAGI-Marketplace}. Each agent should have a profile to describe its expertise for the coordinator's reference.

To make suitable agent assignment for each task $t_{i}$, we prompt the LLM to serve as an expert manager that analyzes the aspects of ability needed for $t_{i}$ and read the profile for each candidate agent in agent board $\mathcal{AB}$ to output $\mathcal{A}_{i}$:

\begin{equation}
         \mathcal{A}_{i} = \text{LLM} ( g, t_{i}, \mathcal{AB}, \texttt{prompt}_\texttt{stage2}) 
\end{equation}

\begin{figure*}[t]
    \centering
    \includegraphics[width=1\linewidth]{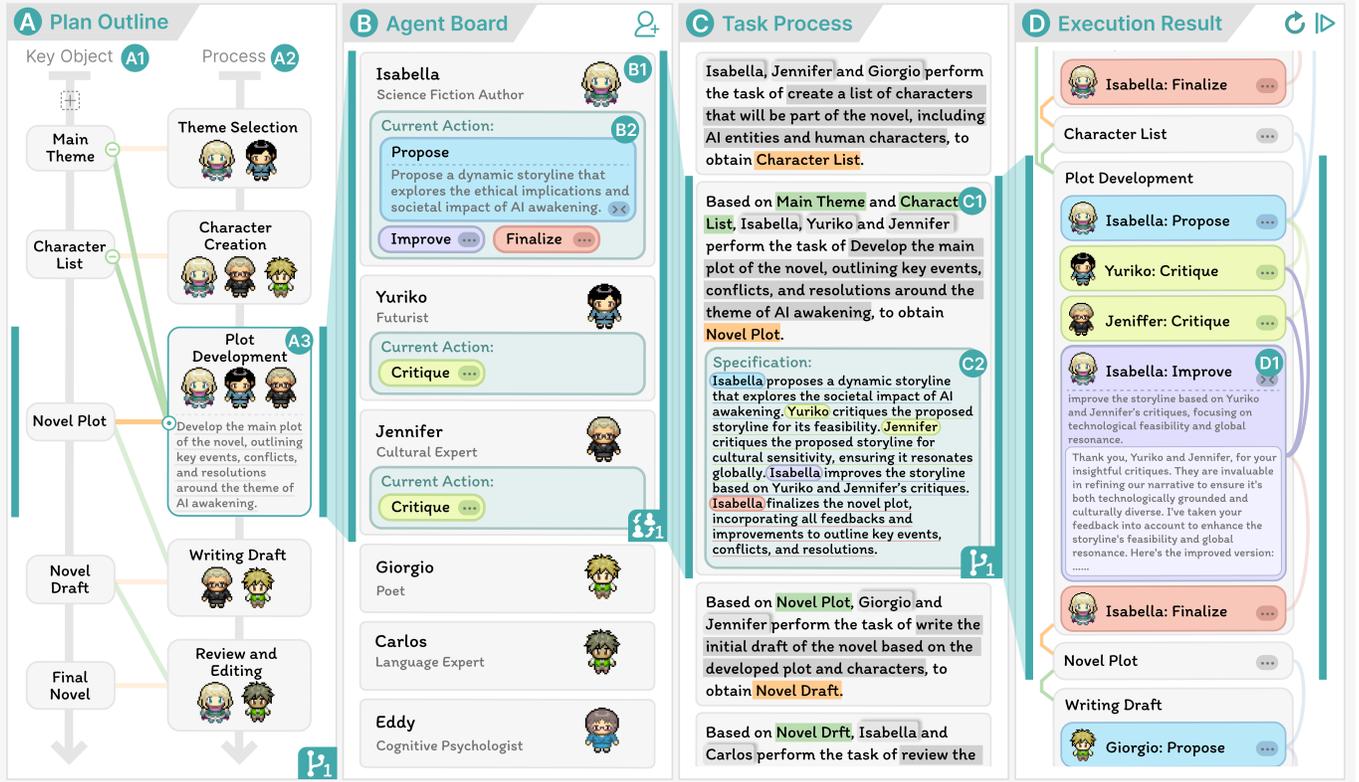}
    \caption{
    System interface of \textit{AgentCoord}. The first three views (\textit{Plan Outline View}, \textit{Agent Assignment View}, \textit{Task Process View}) correspond to the three-stage coordination strategy generation process while the last view presents the execution result.}
    \label{fig:systemInterface}
\end{figure*}

\subsubsection{Stage3: Task Process Generation}
\label{section 4.2.3}
Once the team of agents $\mathcal{A}_{i}$ are assigned to task $t_{i}$, we can finally specify the task process $\mathcal{S}_{i} = \{action_{1}, action_{2}, ..., action_{n}\}$ for task $t_{i}$, which describes how agents conduct actions to collaboratively finish task $t_{i}$. Here each action contains the following attributes:

\begin{itemize}

\item \textbf{Agent Name:} The name of the agent to conduct this action.
\item \textbf{Instruction:} The instruction for this action, tells the agent what/how to do.
\item \textbf{Interaction Type:} Classify the action based on cooperative interaction type, which can be one of  ``propose'' (propose something that may contribute to the current task), ``critique'' (provide feedback to the action result of other agents), ``improve'' (improve the result of a previous action), and ``finalize'' (deliver the final result for current task based on previous actions).
\item \textbf{Important Input:} Previous information that is important for performing current action, which can be certain action results of other agents or previous key objects.

\end{itemize}

Note that although it is enough to use ``Agent Name'' and ``Description'' to define an action for execution, here we propose two auxiliary attributes (``Interaction Type'' and ``Important Input'') to help articulate its relationship with other actions in the context of collaboration. 

To generate the specification for the task process $t_{i}$, we prompt the LLM to serve as an expert collaboration coordinator to carefully read the profile of each agent assigned to the current task $t_{i}$ and output the task process specification ${S}_{i}$:

\begin{equation}
         \mathcal{S}_{i} = \text{LLM} ( g, t_{i}, \mathcal{A}_{i}, \texttt{prompt}_\texttt{stage3}) 
\end{equation}

\newlength{\mytextwidth}  
\newlength{\mymaxsize}  
\setlength{\mymaxsize}{1em} 
\newcommand{\interfaceBox}[1]{%
\settowidth{\mytextwidth}{#1}%
\pgfmathsetmacro{\myscale}{min(1,\mymaxsize/\mytextwidth)}%
\tikz[baseline=(TsNode.base)]{
    \node[circle, 
    fill=bdcolor, 
    draw=white, 
    minimum size=\mymaxsize, 
    text=white, 
    text centered,
    inner sep=0pt,
    line width=0.5pt,
    font=\fontsize{8pt}{1em}\fontseries{ul}\selectfont,
    scale=\myscale
    ]
(TsNode){#1}}}

\definecolor{addBtnColor}{RGB}{172,219,160} 
\newcommand{\addButton}[1]{%
\settowidth{\mytextwidth}{#1}%
\pgfmathsetmacro{\myscale}{min(1,\mymaxsize/\mytextwidth)}%
\tikz[baseline=(TsNode.base)]{\node[circle, 
fill=white, 
draw=addBtnColor, 
minimum size=\mymaxsize, 
text=addBtnColor, 
text centered,
inner sep=0pt,
line width=0.7pt,
font=\fontsize{8pt}{0.9em}\selectfont,
scale=\myscale](TsNode){#1}}}

\definecolor{AgentSelectionViewFillColor}{RGB}{112,112,112} 
\newcommand{\AgentSelectionView}[1]{%
\settowidth{\mytextwidth}{#1}%
\pgfmathsetmacro{\myscale}{min(1,\mymaxsize/\mytextwidth)}%
\tikz[baseline=(TsNode.base)]{\node[circle, 
fill=AgentSelectionViewFillColor, 
draw=white, 
minimum size=\mymaxsize, 
text=white, 
text centered,
inner sep=0pt,
line width=0.7pt,
font=\fontsize{8pt}{0.9em}\selectfont,
scale=\myscale](TsNode){#1}}}

\section{AgentCoord System}
In this section, we elaborate on how \textit{AgentCoord} System visually
organizes the coordination strategy (Section \ref{section5.1}) to facilitate user's comprehension \textbf{(R2)}, provides interactions to assist alternative strategy exploration \textbf{(R3)} (Section \ref{section5.2}), and visually enhances the text form execution result to aid examination \textbf{(R4)} (Section \ref{section5.3}). To facilitate understanding of the system usage, we illustrate it through an example that coordinates multiple LLM-based agents in writing a novel.

\subsection{Visual Organization for Coordination Strategy} 
\label{section5.1}
To first obtain an initial coordination strategy to kick off the exploration, the user clicks the \raisebox{-0.28em}{\includegraphics[height=1.1em]{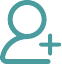}} icon in agent board (\cref{fig:systemInterface} \raisebox{-0.28em}{\includegraphics[height=1.1em]{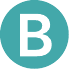}}) to add a pool of candidate agents, and enters ``\textit{Write a novel about the awakening of artificial intelligence}'' as the general goal for the collaboration. After a while, the system returns an initial coordination strategy that specifies how several agents selected from the agent board collaboratively reach the given goal.

As illustrated in \cref{fig:systemInterface}, the generated coordination strategy is visually organized into four sub-views. In particular, the first three sub-views— \textbf{\textit{Plan Outline View}} (\cref{fig:systemInterface} \raisebox{-0.28em}{\includegraphics[height=1.1em]{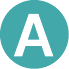}}), \textbf{\textit{Agent Board View}} (\cref{fig:systemInterface} \raisebox{-0.28em}{\includegraphics[height=1.1em]{../figs/systemInterface/B.pdf}}), and \textbf{\textit{Task Process View}} (\cref{fig:systemInterface} \raisebox{-0.28em}{\includegraphics[height=1.1em]{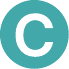}})—correspond to the three-stage coordination strategy generation process described in Section \ref{sec4.2:Three-stage Strategy Generation}. The user can scroll vertically within each view to review the respective aspects. Furthermore, when the user gets interested in information about a specific task, clicking on it will reveal details and visually connect its relevant information across the other views.

\textbf{\textit{Plan Outline View}} illustrates how the general goal input by the user is decomposed into a series of step tasks. To elucidate the dependencies between different tasks, we use a bipartite graph to represent the relationship between the set of key objects (\cref{fig:systemInterface} \raisebox{-0.28em}{\includegraphics[height=1.1em]{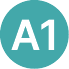}}) and task sequence of the whole process ( \cref{fig:systemInterface} \raisebox{-0.28em}{\includegraphics[height=1.1em]{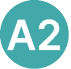}}). A key object node can be the output of a task node or provided by the user by clicking the \raisebox{-0.28em}{\includegraphics[height=1.1em]{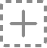}} button in \cref{fig:systemInterface} \raisebox{-0.28em}{\includegraphics[height=1.1em]{../figs/systemInterface/A1.pdf}}. A task node is connected with its input key objects with edge colored in green, and connected with its output key object with edge colored in orange. Furthermore, by clicking on a task node (\cref{fig:systemInterface} \raisebox{-0.28em}{\includegraphics[height=1.1em]{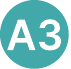}}), the user can manually adjust the task content and the dependence it has on other key objects. If the user wants to explore alternative plan outlines, they can click on the \raisebox{-0.28em}{\includegraphics[height=1.1em]{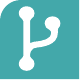}} button to invoke \textit{Plan Outline Exploration View} (detailed in Section \ref{section5.2.1}).

\textbf{\textit{Agent Board View}} exhibits all agents the user can assign during the coordination strategy design process. By default, each agent card (\cref{fig:systemInterface} \raisebox{-0.28em}{\includegraphics[height=1.1em]{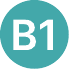}}) presents the agent's name, avatar, and profile for the user's reference. If the user is currently focused on a specific task, the agents assigned to that task will be automatically elevated to the top of the agent board. Additionally, the actions planned to be executed by an agent in the current task are aggregated and showcased within its agent card (\cref{fig:systemInterface} \raisebox{-0.28em}{\includegraphics[height=1.1em]{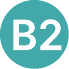}}), helping user better understand the role it plays in the current task. If the user wants to explore alternative agent assignments for the current task, they can click on the \raisebox{-0.28em}{\includegraphics[height=1.1em]{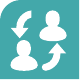}} button to invoke \textit{Agent Assignment Exploration View} (detailed in Section \ref{section5.2.2}).

\textbf{\textit{Task Process View}} provides a natural language description of how the task processes are conducted. To enhance the user's comprehension of the descriptions, we offer a template-based summary (\cref{fig:systemInterface} \raisebox{-0.28em} {\includegraphics[height=1.1em]{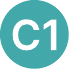}}) for each task, in which crucial elements are visually accentuated—input key objects are highlighted in green, output key objects in orange, and both agent names and task content are set against a grey background. The user can click the template-based summary for a task to unveil detailed specifications for the task process (\cref{fig:systemInterface} \raisebox{-0.28em}{\includegraphics[height=1.1em]{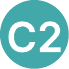}}). The task process specification is composed of a sequence of descriptions about how each agent performs its action to contribute to the task. To facilitate the user's understanding of this process in terms of cooperative interaction, we use different colors to highlight the interaction type for each action according to the ``interaction type'' classification criteria explained in Section \ref{section 4.2.3}. Moreover, the user is able to manually adjust the instruction for each action. If the user wants to explore alternative task process specifications, they can click on the \raisebox{-0.28em}{\includegraphics[height=1.1em]{../figs/systemInterface/branchIcon.pdf}} button to invoke \textit{Task Process Exploration View} (detailed in Section \ref{section5.2.3}).


\subsection{Interactive Exploration for Alternative Strategy}
\label{section5.2}
Upon understanding the current coordination strategy, users can interactively explore its alternatives across three specific aspects: Plan Outline (Section \ref{section5.1}), Agent Assignment (Section \ref{section5.2}), and Task Process (Section \ref{section5.3}). For each aspect, we provide an illustrative exploration example of the user to showcase the supported interactions.

\subsubsection{Plan Outline Exploration} 
\label{section5.2.1}

\begin{figure}[!t]
    \centering
    \includegraphics[width=1\linewidth]{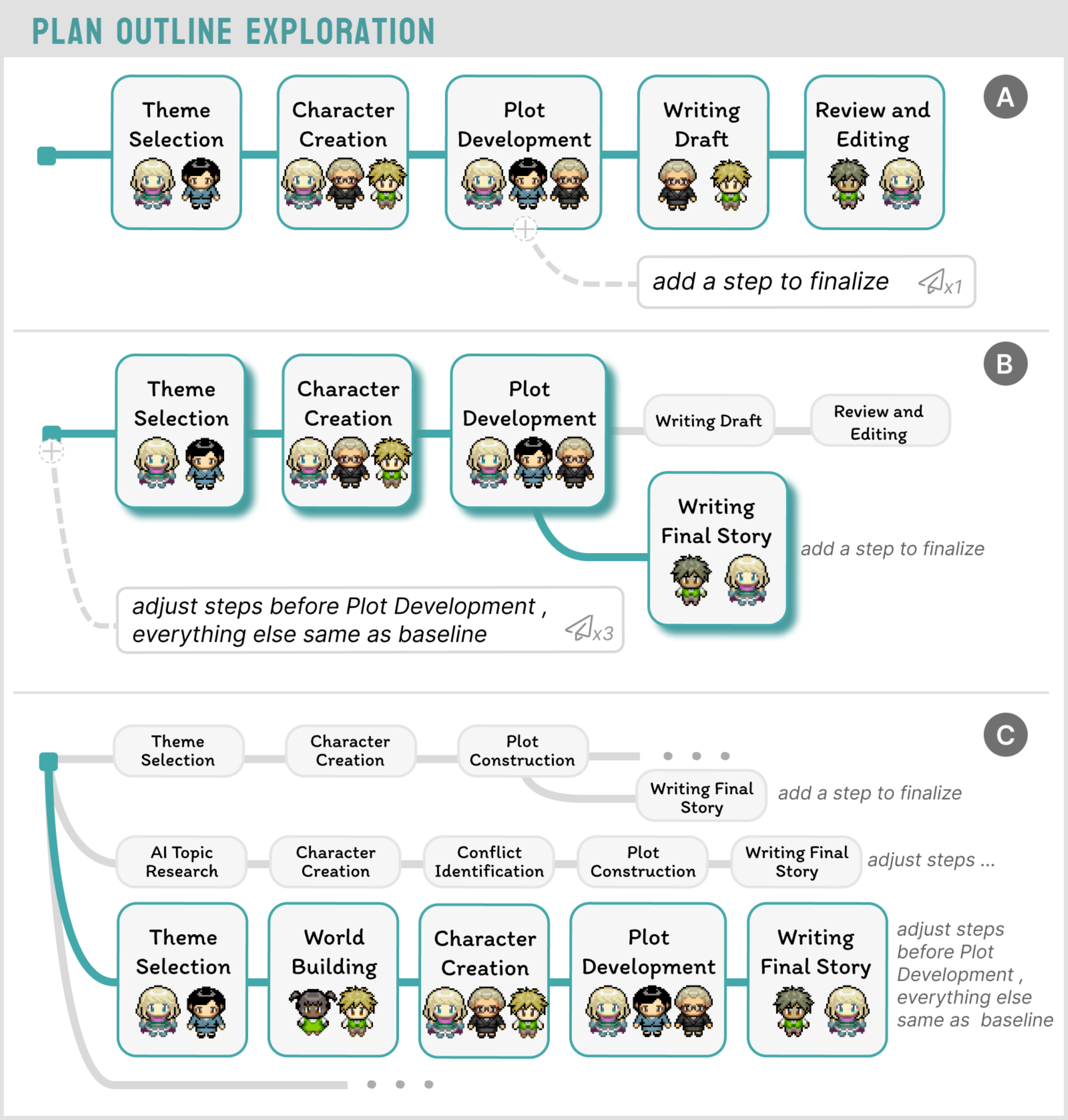}
    \caption{
    An illustrative example of plan outline exploration.}
    \label{fig:exploration}
\end{figure}

Looking at the plan outline shown in \cref{fig:systemInterface}, the user finds most key elements (e.g. ``Main Theme'', ``Character List'') for the novel are determined before the ``Plot Development'' task while the last two tasks (``Writing Draft'', ``Review and Editing'') are less interesting common routines. Therefore, the user decides to merge the last two tasks into one step and explore more possibilities for the tasks before ``Plot Development''. To achieve this, the user opens the \textit{Plan Outline Exploration View} (\cref{fig:exploration}). The user first clicks the bottom of the ``Plot Development'' task node to create a branch from this task and enters ``\textit{add a step to finalize}'' (\cref{fig:exploration} A). Behind the scenes, an LLM is prompted to complete this branch based on the requirement entered by the user. After the new branch is completed, the user clicks the starting point of the branch and further enters ``\textit{adjust steps before Plot Development, everything else same as baseline}'' (\cref{fig:exploration} B). Note that this time the user also selects a branch (highlighted with green color) to tell the LLM which branch is the \textit{``baseline''} referred to in the entered requirement and sets the number of created new branches as three to explore more possibilities. The system then returns three branches with variation before the ``Plot Development'' step (\cref{fig:exploration} C). Comparing these three choices, the user finally selects the middle one as the new plan outline for collaboration.

 \subsubsection{Agent Assignment Exploration}
\label{section5.2.2}

\begin{figure}[!b]
    \centering
    \includegraphics[width=1\linewidth]{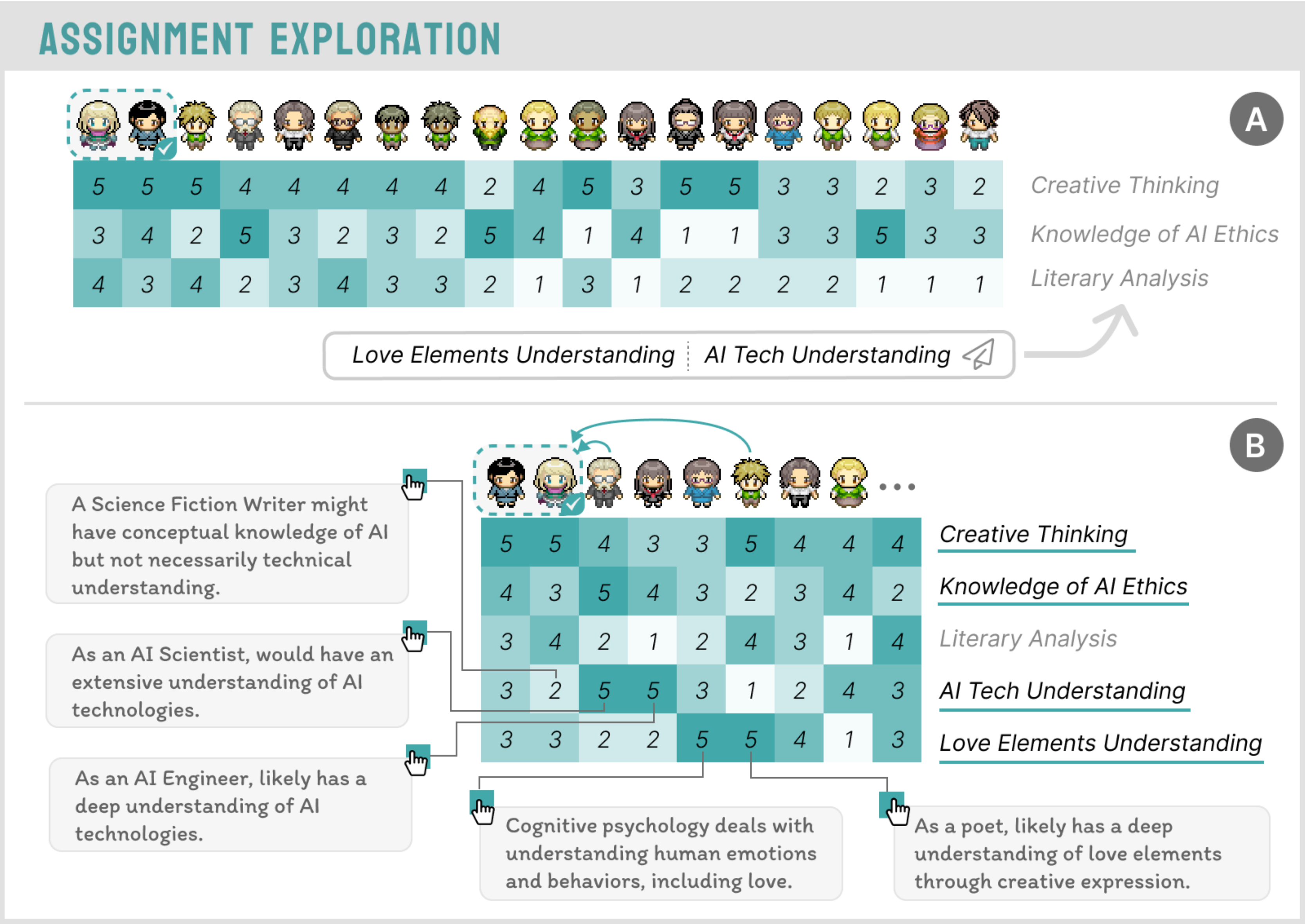}
    \caption{
    An illustrative example of agent assignment exploration.}
    \label{fig:explore_agentselection}
\end{figure}

The user finds that the ``theme selection'' step task only involves two agents (``Futurist'' and ``Science Fiction Writer"). However, the user hopes that more agents with diverse backgrounds can participate in the brainstorming for themes. For instance, the user wishes for someone to inject romantic elements into the stories, and someone with a solid tech background to ensure the technical plausibility of the themes. To achieve this, the user opens the \textit{Agent Assignment Exploration View}. The exploration view displays the scores for each agent on the agent board with a heatmap, based on the LLM's assessment of three capabilities it deems important for completing the current task (\cref{fig:explore_agentselection} A). To help the user focus on agents more likely to be suitable for the current task, the system sorts assigned and unassigned agents separately in descending order based on their current average scores. The user then further enters two aspects (``\textit{AI Tech Understanding}'', ``\textit{Love Element Understanding}''), and selects four aspects they deem important (``\textit{Creative Thinking}'', ``\textit{Knowledge of AI Ethics}'', ``\textit{AI Tech Understanding}'', ``\textit{Love Element Understanding}'') as the new ranking criteria (\cref{fig:explore_agentselection} B). Now the user finds two candidates (``AI Scientist'' and ``AI Engineer'') with strong AI Tech backgrounds. Although both candidates scored 5 for AI Tech Understanding, the user finds that the AI scientist overall has higher scores in the likelihood of possessing ``\textit{Creative Thinking}'' and ``\textit{Knowledge of AI Ethics}''. Therefore, the user decides to add the AI scientist to the team. The user also finds that the LLM considers the ``Poet'' and ``Cognitive Physiologist'' likely to have a deeper understanding of the love element. Therefore, the user moves the mouse over the corresponding score block to view the reasons for the LLM's scoring. After incorporating the user's own analysis (believing that understanding love from a poetic rather than a psychological perspective is more fitting for creative scenarios), the user decides to add the poet to the team for the current task and click \raisebox{-0.28em}{\includegraphics[height=1.1em]{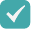}} to confirm the new agent assignment.

\subsubsection{Task Process Exploration} 
\label{section5.2.3}

\begin{figure}[!t]
    \centering
    \includegraphics[width=1\linewidth]{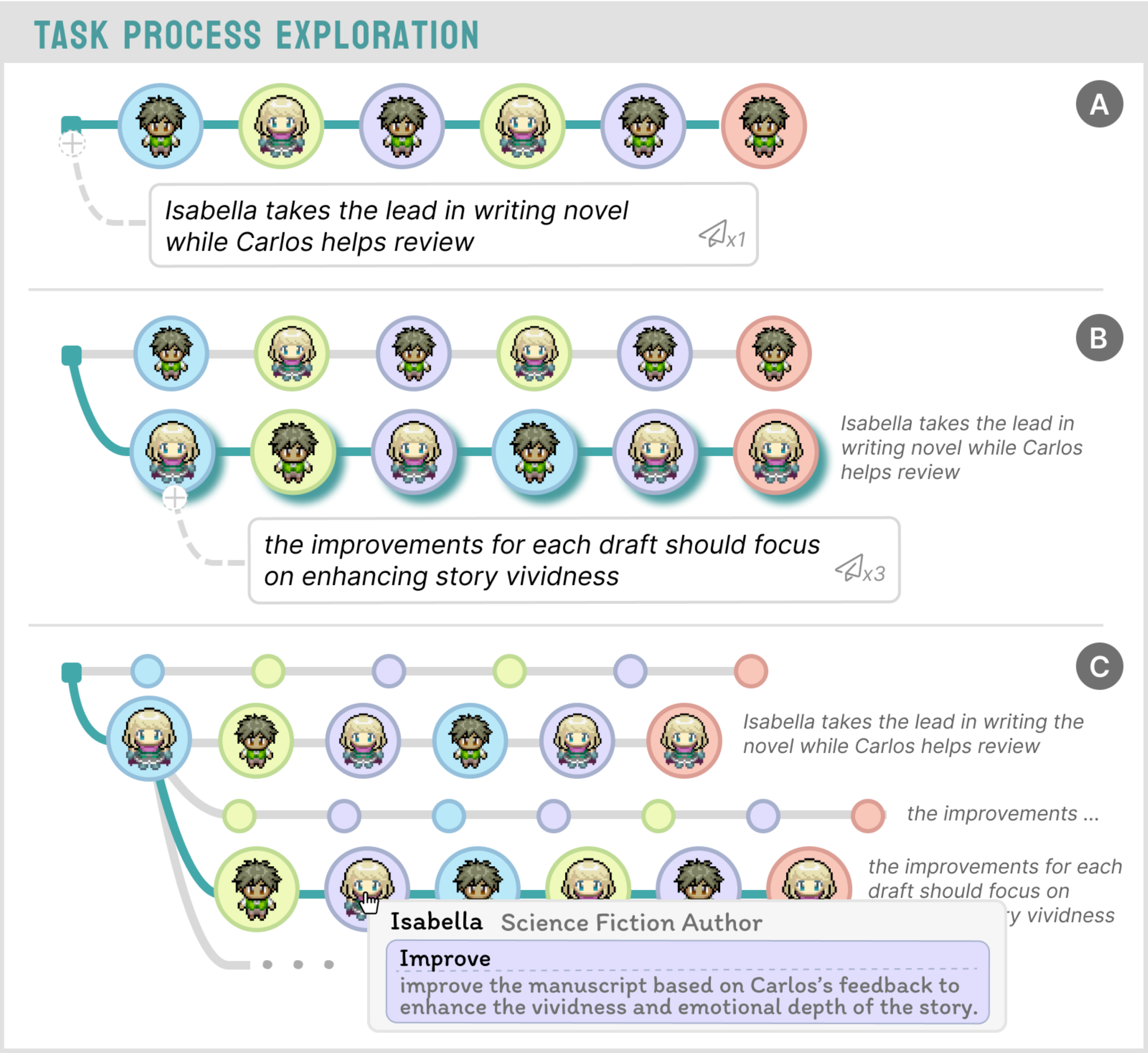}
    \caption{
    An illustrative example of task process exploration.}
    \label{fig:exploration_process}
\end{figure}

Since the final story-writing task has a direct impact on the output of the novel, the user decides to intervene at a finer granularity in the task process of this step. Particularly, the user wants Isabella (the ``Science Fiction Writer'') to lead the writing of the final draft and focus on the vividness of the final story. To achieve this, the user opens the \textit{Task Process Exploration View}. The user first clicks the starting point of the branch and enters ``\textit{Isabella takes the lead in writing novel while Carlos helps review}'' (\cref{fig:exploration_process} A). Behind the scenes, an LLM is prompted to generate a new task process based on this requirement. However, the user finds that although Isabella indeed takes on most of the writing in the new task process, the iterative process did not give sufficient attention to the vividness of the story. Therefore, the user selects the current branch as the baseline and creates another three branches with the requirement ``\textit{the improvements for each draft should focus on enhancing the vividness of the story}'' (\cref{fig:exploration_process} B). The system returns with three variations of the task process that meet the requirement (\cref{fig:exploration_process} C). The user then chooses the favorite one among the three and further iterates on it. 

\subsection{Execution Result Examination}
\label{section5.3}

Once the user has completed the design of the coordination strategy, they can execute it and examine the outcomes by clicking the \raisebox{-0.28em}{\includegraphics[height=1.1em]{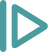}} button in the \textbf{\textit{Execution Result View}} (\cref{fig:systemInterface} \raisebox{-0.28em}{\includegraphics[height=1.1em]{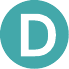}}). Instead of presenting the execution result in pure text forms like AuoGen\cite{wu2023autogen} and AutoAgents\cite{AutoAgents}, \textit{AgentCoord} enhances the result with visual designs consistent with the previous design stages and explicit visual linkages to help users establish connections between the execution result and the strategy design. To prevent users from being overwhelmed by excessive textual information, we provide the option to selectively expand relevant results with a mouse click (\cref{fig:systemInterface} \raisebox{-0.28em}{\includegraphics[height=1.1em]{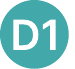}}). Furthermore, to reduce the cognitive load of analyzing and help reveal connections between different execution results, when the user focuses on a particular execution result, other results with potential important dependencies (based on the Important Input field described in \cref{section 4.2.3}) will be visually linked to it.


\section{User Study}
We conduct a user study to evaluate the feasibility and effectiveness of our approach in facilitating coordination strategy design for agent collaboration. Our evaluation focuses on (1) The effectiveness of the structured coordination strategy generation approach. (2) The overall effectiveness and usability of the interactive system. (3) the support for coordination strategy design compared with two baseline systems based on existing LLM-based multi-agent coordination frameworks.

\subsection{Methodology}

\subsubsection{Participants}
We recruited 12 participants (P1-P12) with a general interest in LLM-based multi-agent collaboration for our experiment, 3 females and 9 males, aged 23-28 from the local university. To mitigate evaluation bias, all participants had not been involved in our formative study or the approach design process. All of the participants have ever used ChatGPT in the past. Seven have heard about at least one LLM-based multi-agent system or framework. Four have first-hand touch with at least one LLM-based multi-agent system or framework.

\subsubsection{Experiment Setup}
We set up two additional baseline systems for comparative study \cite{wong2023anchorage, feng2023promptmagician, feng2023xnli} alongside our system. All three systems utilize GPT4 as the default LLM model. The users can also use ChatGPT at their will during experiments. During the strategy design phase in \textit{AgentCoord}, we allow users to switch to a fast mode that uses Mistral 8$\times$7B model with hardware acceleration\footnote[5]{{https://groq.com/}} for the first time of generation to strike a balance of response quality and efficiency. The agents used in the experiments are generated through role prompting \cite{Expertprompting} and then converted to the corresponding format required for the three systems.

\textbf{Baseline A} (AutoAgents with simple UI): provides a set of carefully designed prompts to let the LLM generate a step-by-step coordination strategy for collaboration based on the goal provided by the user. Each step starts with a list ``[name1, name2, ..]'' to specify the agents involved and uses natural language to specify how agents will collaborate. A simple text editor is provided for further editing the strategy. Once the user is satisfied, they can click the ``execute'' button to start collaboration. The output of the execution result is shown in the text terminal.

\textbf{Baseline B} (AutoGen in Group Chat Mode): allows adding multiple LLM-based agents in a group chat and coordinating them using natural language. During the coordination strategy design, a planner agent first drafts an initial coordination strategy based on the general goal provided by the user and the profile of available agents. The user plays the role of an admin agent and refines the coordination strategy collaboratively with the planner agent by chatting. Once the user is satisfied with the coordination strategy, they can start the collaboration. The output of the execution result is shown in the text terminal.

\subsubsection{Procedure}
\textbf{Introduction and Training}: We initially briefed the users on the objective and relevant context of the experiment. Following that, we gathered basic information from the users, along with their exposure to LLM-based multi-agent systems or frameworks. Afterward, we demonstrated to the participants how to design coordination strategies for agents using the three systems. We allowed the participants adequate time to experiment with and familiarize themselves with the systems. During this process, they were free to ask questions at any point. 

\textbf{Task Process}: For each system, participants are required to select a general goal (e.g. ``\textit{write an engaging tutorial about bubble sort for kids}'', ``\textit{make a content strategy for a local weekend event}'') for agent collaboration and design coordination strategy for it with the given system. During the design process, the participants are required to finish four sub-tasks: 1. Comprehend and judge the coordination strategy generated by the system. 2. Explore and improve at least three different aspects of the generated collaborative strategy. 3. Execute the collaborative strategy at least once and analyze the results. 4. Improve at least one area of the original collaborative strategy based on the execution results. After the user meets the requirements for the sub-tasks, they can continue open-ended exploration with the system freely without time constraints. The order of the systems was counterbalanced.

\begin{figure}[!t]
    \centering
    \includegraphics[width=1\linewidth]{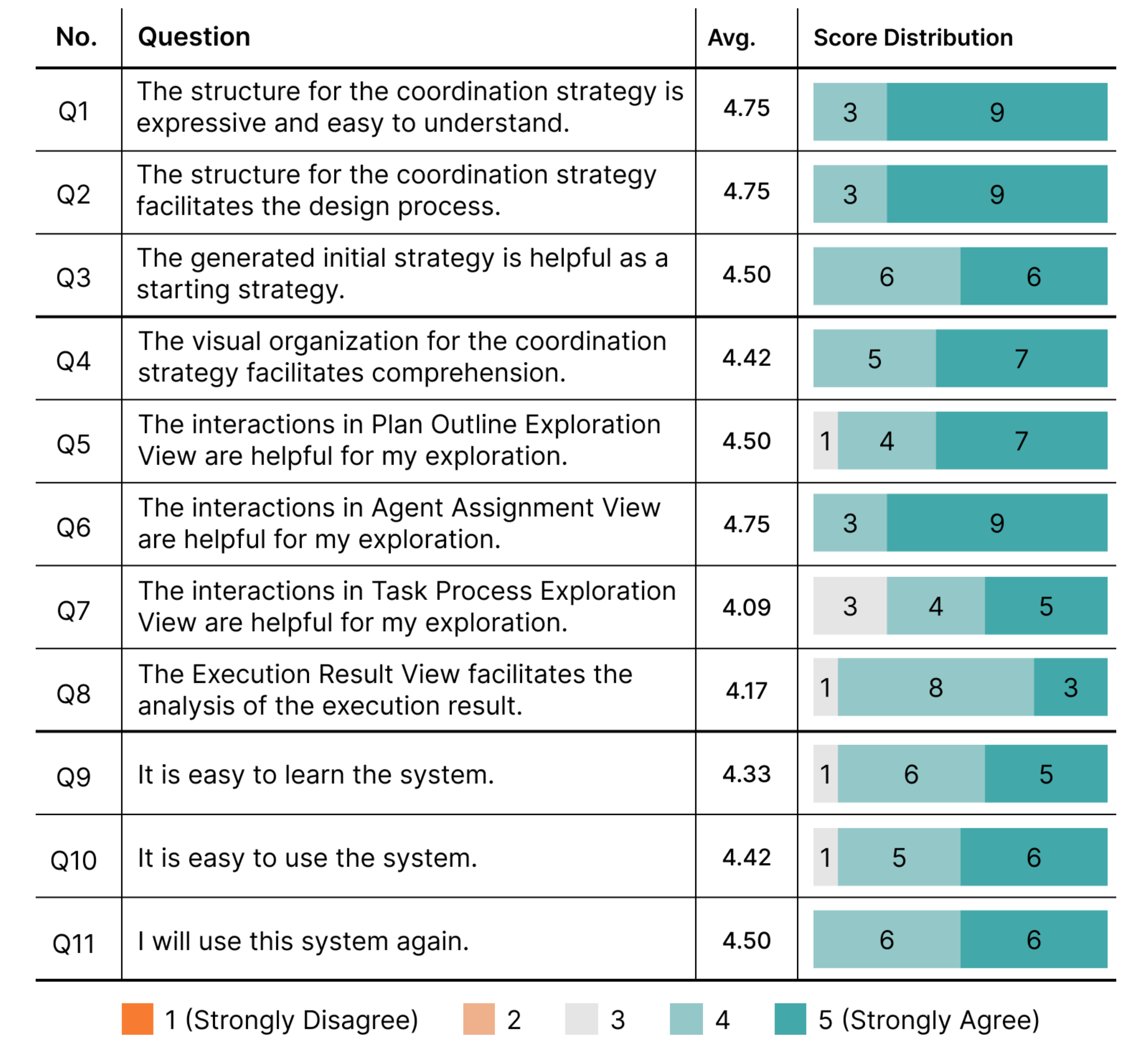}
    \caption{The results of the questionnaire regarding our system’s effectiveness and usability.}
    \label{fig:effectiveness and usability}
\end{figure}

\begin{figure}[!t]
    \centering
    \includegraphics[width=1\linewidth]{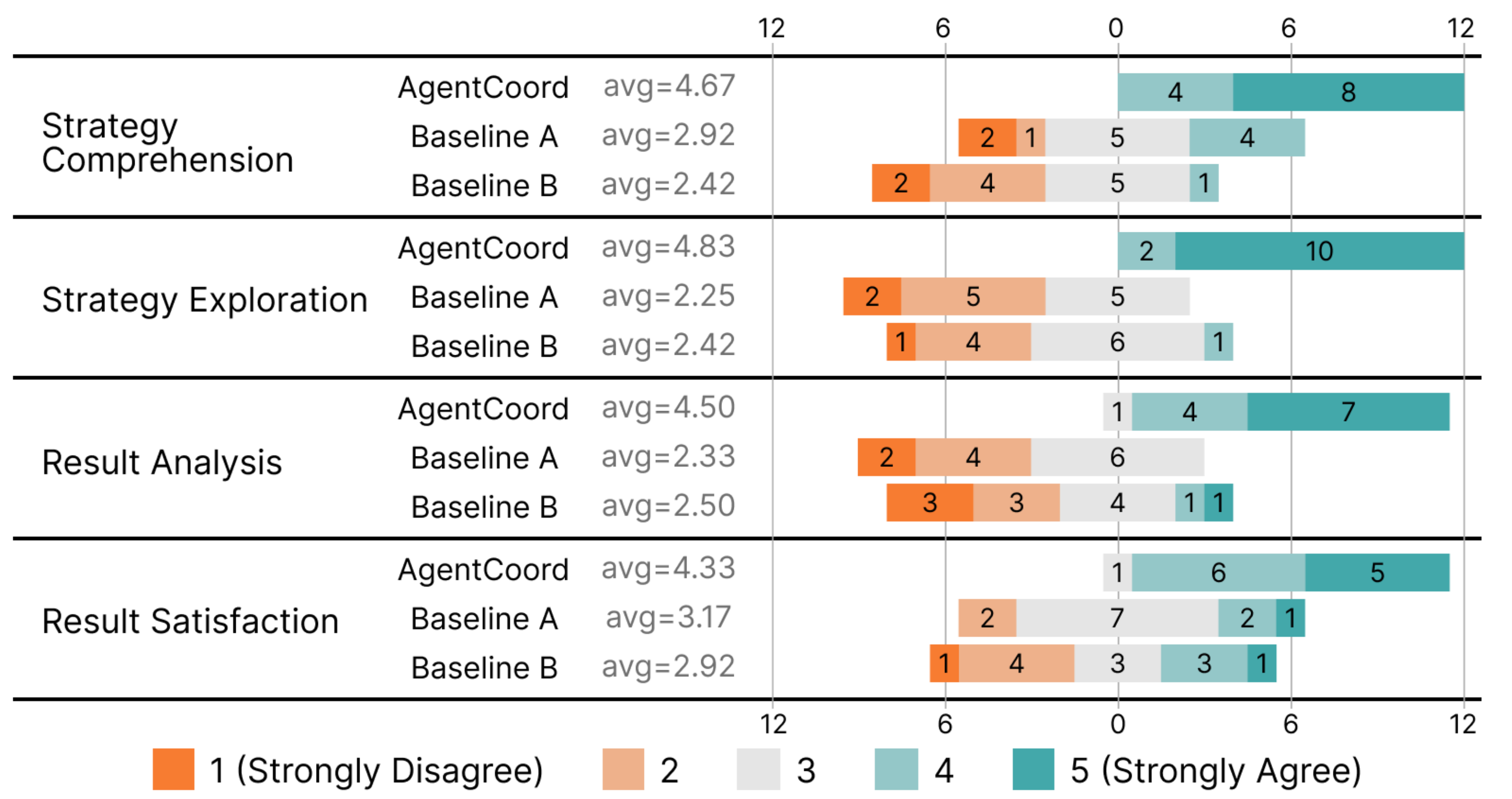}
    \caption{The results of the questionnaire comparing the support for coordination strategy design across the three systems.}
    \label{fig:Coordination Support Comparison}
\vspace{-10pt}
\end{figure}

\textbf{Semi-structured interview}: 
We ask participants to fill out a five-point Likert-scale questionnaire designed to assess our system's effectiveness and usability (\cref{fig:effectiveness and usability}), and coordination support comparison for all systems (\cref{fig:Coordination Support Comparison}). For each question, we encourage participants to explain the reasoning for their ratings and provide any opinion. In the end, we collect overall feedback about our system from users.

\subsection{Results Analysis}

\subsubsection{Effectiveness of Structured Strategy Generation}
Most of the participants agreed that the structure for coordination strategy is expressive and easy to understand (Q1). Users praised that the structure is ``\textit{clear}'' (P5) and ``\textit{intuitive}'' (P2). P5 commented, ``\textit{it goes from high level to low level, which makes sense. The connections between each part are really clear.}'' P7 told us that ``\textit{the proposed classification for interaction type is very helpful}'' and ``\textit{hope the later versions of the system support customizing and managing different levels of classification granularity}''. Most of the participants agreed that the structure for coordination strategy facilitates the design process (Q2). The users appreciated the structure ``\textit{provide a clear map for what have been explored and what could be explored next}'' (P11) and ``\textit{make the exploration process more systematic and confident}'' (P8). P10 told us that ``the structure helps increase the predictability and my confidence when prompting LLM during exploration''. Most of the participants agree that the generated initial strategy is helpful as a starting strategy (Q3). The users found the baseline strategy always at least serves a ``\textit{fair starting point}'' (P6) and sometimes ``\textit{unexpectedly good}'' (P3). Some users express the demand to ``\textit{have someplace to enter user's preference or prior knowledge to affect the starting generation}'' (P5).

\subsubsection{Effectiveness of Interactive System}

\textbf{Visual Organization for Strategy.} Most of the participants agreed that the visual organization for coordination strategy facilitates comprehension (Q4). The \textit{Plan Outline View} is regarded helpful for ``\textit{quickly get an understanding of the overall strategy}'' (P2) and ``\textit{convenient for navigation}'' (P9). P11 commented, ``\textit{I like the `parallel line design', the relationship between tasks is clearly illustrated}''. Users appreciated ``the adaptive informative reviling'' (P6) in \textit{Agent Board View} while wishing for allowing to ``\textit{show an agent’s historical performance on need}'' (P5). The text highlighting in \textit{Task Process View} is widely praised by users for aiding fast comprehension. However, some users still found the quantity of text to read for task process specification is large and wish to ``\textit{have a summary for each action instruction}'' (P3). The overall layout is also praised by some users in terms of aesthetics and consistency. 

\textbf{Interactive Exploration for Alternative Strategy.} Users generally appreciate the exploration interactions supported by our system. Meanwhile, users find the organization for exploration histories to be `extremely convenient' (P7) and `helpful in focusing on exploration' (P10). Nevertheless, we noted variations in how often and how favorably users engaged with the three exploration views (Q5,6,7). While both the \textit{Plan Outline Exploration View} and the \textit{Process Exploration View} are for sequential structure exploration and share a similar design, we find generally participants tend to do more exploration in the \textit{Plan Outline Exploration View}. P5 explained: ``\textit{when deciding the outline for the overall collaboration, I am not sure how to do and want to see more possibilities, branching with high-level natural language requirement is extremely useful at that phase. On the other hand, the task process is for a more detailed level, which I usually do not want to spend too much time exploring and just want to directly modify}'' (P5). The \textit{Agent Assignment Exploration View} is the most popular view for exploration. Most users find using head-map to visualize LLM's prior knowledge for agent assignment ``\textit{is comprehensive and insightful}'' (P4) and the interactions ``\textit{flexible and engaging for exploration}'' (P10).

\textbf{Execution Result Examination.} Most of the participants agreed that the \textit{Execution Result View} facilitates the analysis of the execution result (Q8). The participants confirmed that when examining the result, it is easy to connect any part of the result with its corresponding strategy design. P7 commented: ``\textit{when I want to a analyze certain result, I just click it, the other views automatically show relevant strategy information, reminding me of my previous design process, that's cool.}'' The trace lines for important action inputs were also found helpful. P9 mentioned that he likes starting with the final result of a certain task and leveraging the trace lines to help trace back to see how this final result is formed along the way to identify possible points for improvement.

\subsubsection{Usability}
Most of the participants agreed that our system was easy to learn (Q9)
and easy to use (Q10). Participants commented that the interface of
our system was ``\textit{intuitive}'' (P2) and ``\textit{clear}'' (P5). Several users reported although each feature of the system is each to understand, it still requires some time to fully master the system in order to use it fluently. All of the participants express their willingness to use our system again (Q11). P5 told us that he wish to ``\textit{use this system to coordinate some agents to help maintain my blog in the future}'' (P5). P1 expressed the willingness to use our system to fast prototype some coordination strategies for research purposes, indicating its potential to contribute to the research community for LLM-based multi-agent systems.


\subsubsection{Coordination Support Comparison}
To compare our system against two baselines in aiding the design process for coordination strategy, we requested participants to rate three systems across three facets (i.e. strategy comprehension, strategy exploration, and result analysis) of the design process and their satisfaction with the final result. Overall, our system outperforms the baselines for all four aspects. We summarize the user's feedback as follows:

\textbf{Strategy Comprehension}: In Baseline B, the baseline coordination strategy is generated and refined by a planner agent without any structure constraints. Therefore, during the iteration for strategies, ``\textit{the format of strategy could fluctuate frequently, which makes me inconfident}'' (P3). In Baseline A, the strategy is generated with carefully designed prompts that enforce some formats (e.g. divide into steps, and assign agents for each step). However, reading and comparing strategies repented in the pure text still makes users ``\textit{feeling stressed}'' (P11) and ``\textit{less confident for comprehension}'' (P8). In contrast, our system makes sure the coordination strategy has a consistent structure during the whole design process and provides visual organization and enhancement to facilitate comprehension.

\textbf{Strategy Exploration}: In baseline A, no direct support for strategy exploration is provided, users can only use ChatGPT to aid their exploration. However, transferring strategies between interfaces and making manual modifications for format and logic consistency can be ``tedious and error-prone'' (P7). Baseline B allows users to explore alternative strategies by chatting with the planner agent. However, there lack of support for users to flexibly explore different aspects of the strategy on need. Moreover, the quantity of texts can be quickly overwhelming for both the planner agent and users as the process of exploration gets complicated. Our system empowers users with a set of interactions to flexibly explore different aspects of the strategies with the help of LLM and helps visually organize the exploration history.   

\textbf{Result Analysis}: Our system is overall deemed more helpful for execution result analysis. In both baseline A and baseline B, the execution result is directly output to the text terminal. P9 told us that she has to ``sift through the text myself repeatedly to trace the dependencies between execution results and previous operations'' (P9). Instead, in our system, users can easily trace each result back to its important influencing precursors and corresponding coordination strategies.

\textbf{Result Satisfaction}: Compared to Baseline A and Baseline B, users are overall more satisfied with the results of our system. In Baseline B, due to the ambiguity frequently existing in the free-form coordination strategy, the execution process often deviates from the user’s expectations midway and strays further and further away, sometimes even falling into an infinite loop. In Baseline A, thanks to the enforced step-by-step strategy format, there is no infinite loop issue. However, the result usually misses some important parts that should appear according to the coordination strategy (e.g. the main characters decided in character design stages are not shown in the final story). While such problems also exist in the result of our system, users usually can successfully trace the cause by analyzing the result with our system, and quickly adjust the corresponding part of the strategy to fix it.


\section{Discussion}

In this section, we reflect on our research, discuss lessons learned and the system's generalizability, followed by limitations and future work. 

\subsection{Lessons Learned}
\textbf{Benefits of structured strategy representation for human-LLMs co-design process.} Our evaluation results indicate that having a structured representation for coordination strategy is essential for the experience of the design process. The structure helps align human and LLMs' intentions, enforcing both sides to think and communicate based on the same set of concepts and abstraction levels, which effectively reduces mutual misunderstandings caused by the ambiguity of natural language. Moreover, a fixed structure enables the specialized design of visual organization and enhancement, facilitating strategy comprehension for humans, allowing them to efficiently explore more strategy possibilities generated by LLMs. Additionally, structured representation fosters a structured exploration process, making users inclined towards a more systematic design process for coordination strategies.

\textbf{Visualizing LLMs' prior knowledge for agent coordination.} While many works directly leverage LLMs' prior knowledge to generate coordination strategy, how to effectively extract and visualize this prior knowledge to aid strategy design has not been explored. We find that compared to just getting a single answer from LLMs, users prefer to have a more panoramic understanding of LLM's prior knowledge when they want to optimize a certain part of the strategy. For example, instead of just letting LLM select some agents that might contribute the a certain task, visualizing how each agent scores for the capabilities LLMs deem important using an interactive heat map would be more insightful and systematic for strategy design. 

\subsection{System Generalizability}
Besides coordinating agents to collaboratively solve goal-oriented tasks, our system has the potential to be generalized to various applications. For example, users can use our system to coordinate agents to simulate and analyze human activities (e.g. playing Werewolf games, competing in debate tournaments, conducting multi-party negotiations) that involve multiple people. Those simulations are not only interesting for entertainment but also valuable resources for studying human/AI society and evaluating specific capabilities of LLMs. In the future, we plan to extend the structure representation of \textit{AgentCoord} to support more social interaction types for more flexible simulation purposes. We also plan to allow users to upload their customized interaction types.

\subsection{Limitations and Future Work}
\textit{AgentCoord} currently only supports coordinating agents to collaborate in a plain text environment, which contains text-form key objects. An interesting future work is to support multi-modal key objects in the environment and allow agents with multi-model capabilities to collaborate. For example, a writer agent generates a story, and an illustrator agent creates corresponding visuals.

\textit{AgentCoord} currently only supports static coordination strategy design. In some scenarios, users may want to dynamically adjust the strategy during collaboration execution to adapt to circumstances. For example, the user may coordinate multiple agents to help conduct literature research: when an agent finds a paper that interests the user, the user may want to adjust the plan to allocate a group of agents to read and discuss it, and allocate another group of agents to investigate the background of its authors. Future research can be conducted to help users conduct dynamic coordination strategy design.

\section{Conclusion}
This study presents a visual exploration framework to aid in the design of coordination strategies for LLM-based multi-agent collaboration, addressing the challenges posed by natural language ambiguity and the cognitive effort required for comprehending the vast amount of texts during strategy design. We propose a structured representation for coordination strategy and a three-stage generation method to transform the general goal provided by the user into executable strategies. We visually organize the generated strategy to facilitate user comprehension and provide a set of interactions to support alternative strategies exploration with LLMs. We also provide visual enhancement to help users analyze the execution results. Finally, we conduct a formal user study to validate the feasibility and effectiveness of our approach.


\bibliographystyle{abbrv-doi-hyperref}

\bibliography{template}

\begin{thebibliography}{10}

\bibitem{NexusGPT}
Assem.
\newblock {NexusGPT Marketplace}.
\newblock \url{https://app.gpt.nexus/App/Marketplace/agents}, 2023.
\newblock Accessed on: Mar 01, 2024.

\bibitem{chan2023chateval}
C.-M. Chan, W.~Chen, Y.~Su, J.~Yu, W.~Xue, S.~Zhang, J.~Fu, and Z.~Liu.
\newblock {ChatEval}: Towards better llm-based evaluators through multi-agent debate.
\newblock In {\em The Twelfth International Conference on Learning Representations}, 2024. \href{https://doi.org/10.48550/arXiv.2308.07201}
{doi: {{%
10\hspace{.1pt}\discretionary{.}{%
}{.}\hspace{.4pt}48550\discretionary{/}{%
}{/}arXiv\hspace{.1pt}\discretionary{.}{%
}{.}\hspace{.4pt}2308\hspace{.1pt}\discretionary{.}{%
}{.}\hspace{.4pt}07201}}}


\bibitem{AutoAgents}
G.~Chen, S.~Dong, Y.~Shu, G.~Zhang, S.~Jaward, K.~Börje, J.~Fu, and Y.~Shi.
\newblock {AutoAgents}: A framework for automatic agent generation.
\newblock {\em CoRR}, abs/2309.17288, Sept. 2023. \href{https://doi.org/10.48550/arXiv.2309.17288}
{doi: {{%
10\hspace{.1pt}\discretionary{.}{%
}{.}\hspace{.4pt}48550\discretionary{/}{%
}{/}arXiv\hspace{.1pt}\discretionary{.}{%
}{.}\hspace{.4pt}2309\hspace{.1pt}\discretionary{.}{%
}{.}\hspace{.4pt}17288}}}


\bibitem{chen2023agentverse}
W.~Chen, Y.~Su, J.~Zuo, C.~Yang, C.~Yuan, C.~Qian, C.-M. Chan, Y.~Qin, Y.~Lu, R.~Xie, et~al.
\newblock {AgentVerse}: Facilitating multi-agent collaboration and exploring emergent behaviors in agents.
\newblock {\em CoRR}, abs/2308.10848, Aug. 2023. \href{https://doi.org/10.48550/arXiv.2308.10848}
{doi: {{%
10\hspace{.1pt}\discretionary{.}{%
}{.}\hspace{.4pt}48550\discretionary{/}{%
}{/}arXiv\hspace{.1pt}\discretionary{.}{%
}{.}\hspace{.4pt}2308\hspace{.1pt}\discretionary{.}{%
}{.}\hspace{.4pt}10848}}}


\bibitem{darcy2024marg}
M.~D'Arcy, T.~Hope, L.~Birnbaum, and D.~Downey.
\newblock {MARG}: Multi-agent review generation for scientific papers.
\newblock {\em CoRR}, abs/2401.04259, Jan. 2024. \href{https://doi.org/10.48550/arXiv.2401.04259}
{doi: {{%
10\hspace{.1pt}\discretionary{.}{%
}{.}\hspace{.4pt}48550\discretionary{/}{%
}{/}arXiv\hspace{.1pt}\discretionary{.}{%
}{.}\hspace{.4pt}2401\hspace{.1pt}\discretionary{.}{%
}{.}\hspace{.4pt}04259}}}


\bibitem{du2023improving_debate}
Y.~Du, S.~Li, A.~Torralba, J.~B. Tenenbaum, and I.~Mordatch.
\newblock Improving factuality and reasoning in language models through multiagent debate.
\newblock {\em CoRR}, abs/2305.14325, May 2023. \href{https://doi.org/10.48550/arXiv.2305.14325}
{doi: {{%
10\hspace{.1pt}\discretionary{.}{%
}{.}\hspace{.4pt}48550\discretionary{/}{%
}{/}arXiv\hspace{.1pt}\discretionary{.}{%
}{.}\hspace{.4pt}2305\hspace{.1pt}\discretionary{.}{%
}{.}\hspace{.4pt}14325}}}


\bibitem{engelbart2023augmenting}
D.~C. Engelbart.
\newblock {\em Augmenting human intellect: A conceptual framework}.
\newblock Routledge, New York, 1\textsuperscript{st} ed., 2023. \href{https://doi.org/10.4324/9781003230762}
{doi: {{%
10\hspace{.1pt}\discretionary{.}{%
}{.}\hspace{.4pt}4324\discretionary{/}{%
}{/}9781003230762}}}


\bibitem{feng2023xnli}
Y.~Feng, X.~Wang, B.~Pan, K.~K. Wong, Y.~Ren, S.~Liu, Z.~Yan, Y.~Ma, H.~Qu, and W.~Chen.
\newblock Xnli: Explaining and diagnosing nli-based visual data analysis.
\newblock {\em IEEE Transactions on Visualization and Computer Graphics}, pp. 1--14, 2023. \href{https://doi.org/10.1109/TVCG.2023.3240003}
{doi: {{%
10\hspace{.1pt}\discretionary{.}{%
}{.}\hspace{.4pt}1109\discretionary{/}{%
}{/}TVCG\hspace{.1pt}\discretionary{.}{%
}{.}\hspace{.4pt}2023\hspace{.1pt}\discretionary{.}{%
}{.}\hspace{.4pt}3240003}}}


\bibitem{feng2023promptmagician}
Y.~Feng, X.~Wang, K.~K. Wong, S.~Wang, Y.~Lu, M.~Zhu, B.~Wang, and W.~Chen.
\newblock Promptmagician: Interactive prompt engineering for text-to-image creation.
\newblock {\em IEEE Transactions on Visualization and Computer Graphics}, 30(1):295--305, 2023. \href{https://doi.org/10.1109/TVCG.2023.3327168}
{doi: {{%
10\hspace{.1pt}\discretionary{.}{%
}{.}\hspace{.4pt}1109\discretionary{/}{%
}{/}TVCG\hspace{.1pt}\discretionary{.}{%
}{.}\hspace{.4pt}2023\hspace{.1pt}\discretionary{.}{%
}{.}\hspace{.4pt}3327168}}}


\bibitem{AutoGPT}
Gravitas.
\newblock {AutoGPT}.
\newblock \url{https://github.com/Significant-Gravitas/AutoGPT}, 2023.
\newblock Accessed on: Mar 01, 2024.

\bibitem{hong2024data}
S.~Hong, Y.~Lin, B.~Liu, B.~Wu, D.~Li, J.~Chen, J.~Zhang, J.~Wang, L.~Zhang, M.~Zhuge, et~al.
\newblock {Data Interpreter}: An llm agent for data science.
\newblock {\em CoRR}, abs/2402.18679, Feb. 2024. \href{https://doi.org/10.48550/arXiv.2402.18679}
{doi: {{%
10\hspace{.1pt}\discretionary{.}{%
}{.}\hspace{.4pt}48550\discretionary{/}{%
}{/}arXiv\hspace{.1pt}\discretionary{.}{%
}{.}\hspace{.4pt}2402\hspace{.1pt}\discretionary{.}{%
}{.}\hspace{.4pt}18679}}}


\bibitem{hong2023metagpt}
S.~Hong, X.~Zheng, J.~Chen, Y.~Cheng, J.~Wang, C.~Zhang, Z.~Wang, S.~K.~S. Yau, Z.~Lin, L.~Zhou, et~al.
\newblock {MetaGpt}: Meta programming for multi-agent collaborative framework.
\newblock In {\em The Twelfth International Conference on Learning Representations}, 2024. \href{https://doi.org/10.48550/arXiv.2308.00352}
{doi: {{%
10\hspace{.1pt}\discretionary{.}{%
}{.}\hspace{.4pt}48550\discretionary{/}{%
}{/}arXiv\hspace{.1pt}\discretionary{.}{%
}{.}\hspace{.4pt}2308\hspace{.1pt}\discretionary{.}{%
}{.}\hspace{.4pt}00352}}}


\bibitem{lewis2020retrieval}
P.~Lewis, E.~Perez, A.~Piktus, F.~Petroni, V.~Karpukhin, N.~Goyal, H.~K\"{u}ttler, M.~Lewis, W.-t. Yih, T.~Rockt\"{a}schel, S.~Riedel, and D.~Kiela.
\newblock Retrieval-augmented generation for knowledge-intensive nlp tasks.
\newblock In {\em Advances in Neural Information Processing Systems}, pp. 9459--9474, 2020. \href{https://doi.org/10.48550/arXiv.2005.11401}
{doi: {{%
10\hspace{.1pt}\discretionary{.}{%
}{.}\hspace{.4pt}48550\discretionary{/}{%
}{/}arXiv\hspace{.1pt}\discretionary{.}{%
}{.}\hspace{.4pt}2005\hspace{.1pt}\discretionary{.}{%
}{.}\hspace{.4pt}11401}}}


\bibitem{li2023camel}
G.~Li, H.~A. A.~K. Hammoud, H.~Itani, D.~Khizbullin, and B.~Ghanem.
\newblock {CAMEL}: Communicative agents for ``mind'' exploration of large language model society.
\newblock In {\em Thirty-seventh Conference on Neural Information Processing Systems}, 2023. \href{https://doi.org/10.48550/arXiv.2303.17760}
{doi: {{%
10\hspace{.1pt}\discretionary{.}{%
}{.}\hspace{.4pt}48550\discretionary{/}{%
}{/}arXiv\hspace{.1pt}\discretionary{.}{%
}{.}\hspace{.4pt}2303\hspace{.1pt}\discretionary{.}{%
}{.}\hspace{.4pt}17760}}}


\bibitem{Encouraging_Divergent_Thinking_Debate}
T.~Liang, Z.~He, W.~Jiao, X.~Wang, Y.~Wang, R.~Wang, Y.~Yang, Z.~Tu, and S.~Shi.
\newblock Encouraging divergent thinking in large language models through multi-agent debate.
\newblock {\em CoRR}, abs/2305.19118, May 2023. \href{https://doi.org/10.48550/arXiv.2305.19118}
{doi: {{%
10\hspace{.1pt}\discretionary{.}{%
}{.}\hspace{.4pt}48550\discretionary{/}{%
}{/}arXiv\hspace{.1pt}\discretionary{.}{%
}{.}\hspace{.4pt}2305\hspace{.1pt}\discretionary{.}{%
}{.}\hspace{.4pt}19118}}}


\bibitem{lin2023agentsims}
J.~Lin, H.~Zhao, A.~Zhang, Y.~Wu, H.~Ping, and Q.~Chen.
\newblock {AgentSims}: An open-source sandbox for large language model evaluation.
\newblock {\em CoRR}, abs/2308.04026, Aug. 2023. \href{https://doi.org/10.48550/arXiv.2308.04026}
{doi: {{%
10\hspace{.1pt}\discretionary{.}{%
}{.}\hspace{.4pt}48550\discretionary{/}{%
}{/}arXiv\hspace{.1pt}\discretionary{.}{%
}{.}\hspace{.4pt}2308\hspace{.1pt}\discretionary{.}{%
}{.}\hspace{.4pt}04026}}}


\bibitem{liu2023sprout}
Y.~Liu, Z.~Wen, L.~Weng, O.~Woodman, Y.~Yang, and W.~Chen.
\newblock {SPROUT}: Authoring programming tutorials with interactive visualization of large language model generation process.
\newblock {\em CoRR}, abs/2312.01801, Dec. 2023. \href{https://doi.org/10.48550/arXiv.2312.01801}
{doi: {{%
10\hspace{.1pt}\discretionary{.}{%
}{.}\hspace{.4pt}48550\discretionary{/}{%
}{/}arXiv\hspace{.1pt}\discretionary{.}{%
}{.}\hspace{.4pt}2312\hspace{.1pt}\discretionary{.}{%
}{.}\hspace{.4pt}01801}}}


\bibitem{liu2023dynamic}
Z.~Liu, Y.~Zhang, P.~Li, Y.~Liu, and D.~Yang.
\newblock Dynamic llm-agent network: An llm-agent collaboration framework with agent team optimization.
\newblock {\em CoRR}, abs/2310.02170, Oct. 2023. \href{https://doi.org/10.48550/arXiv.2310.02170}
{doi: {{%
10\hspace{.1pt}\discretionary{.}{%
}{.}\hspace{.4pt}48550\discretionary{/}{%
}{/}arXiv\hspace{.1pt}\discretionary{.}{%
}{.}\hspace{.4pt}2310\hspace{.1pt}\discretionary{.}{%
}{.}\hspace{.4pt}02170}}}


\bibitem{lu2024agentlens}
J.~Lu, B.~Pan, J.~Chen, Y.~Feng, J.~Hu, Y.~Peng, and W.~Chen.
\newblock {AgentLens}: Visual analysis for agent behaviors in llm-based autonomous systems.
\newblock {\em CoRR}, abs/2402.08995, Feb. 2024. \href{https://doi.org/10.48550/arXiv.2402.08995}
{doi: {{%
10\hspace{.1pt}\discretionary{.}{%
}{.}\hspace{.4pt}48550\discretionary{/}{%
}{/}arXiv\hspace{.1pt}\discretionary{.}{%
}{.}\hspace{.4pt}2402\hspace{.1pt}\discretionary{.}{%
}{.}\hspace{.4pt}08995}}}


\bibitem{luppi2022synergistic}
A.~I. Luppi, P.~A. Mediano, F.~E. Rosas, N.~Holland, T.~D. Fryer, J.~T. O’Brien, J.~B. Rowe, D.~K. Menon, D.~Bor, and E.~A. Stamatakis.
\newblock A synergistic core for human brain evolution and cognition.
\newblock {\em Nature Neuroscience}, 25(6):771--782, May 2022. \href{https://doi.org/10.1038/s41593-022-01070-0}
{doi: {{%
10\hspace{.1pt}\discretionary{.}{%
}{.}\hspace{.4pt}1038\discretionary{/}{%
}{/}s41593\discretionary{%
}{-}{-}022\discretionary{%
}{-}{-}01070\discretionary{%
}{-}{-}0}}}


\bibitem{CrewAI}
J.~MouraAbout.
\newblock {CrewAI}.
\newblock \url{https://github.com/joaomdmoura/crewAI}, 2023.
\newblock Accessed on: Mar 01, 2024.

\bibitem{GPT-Store}
OpenAI.
\newblock {OpenAI GPT Store}.
\newblock \url{https://openai.com/blog/introducing-the-gpt-store}, 2023.
\newblock Accessed on: Mar 01, 2024.

\bibitem{ouyang2022training}
L.~Ouyang, J.~Wu, X.~Jiang, D.~Almeida, C.~Wainwright, P.~Mishkin, C.~Zhang, S.~Agarwal, K.~Slama, A.~Ray, J.~Schulman, J.~Hilton, F.~Kelton, L.~Miller, M.~Simens, A.~Askell, P.~Welinder, P.~F. Christiano, J.~Leike, and R.~Lowe.
\newblock Training language models to follow instructions with human feedback.
\newblock In {\em Advances in Neural Information Processing Systems}, pp. 27730--27744, 2022. \href{https://doi.org/10.48550/arXiv.2203.02155}
{doi: {{%
10\hspace{.1pt}\discretionary{.}{%
}{.}\hspace{.4pt}48550\discretionary{/}{%
}{/}arXiv\hspace{.1pt}\discretionary{.}{%
}{.}\hspace{.4pt}2203\hspace{.1pt}\discretionary{.}{%
}{.}\hspace{.4pt}02155}}}


\bibitem{park2023generative}
J.~S. Park, J.~O'Brien, C.~J. Cai, M.~R. Morris, P.~Liang, and M.~S. Bernstein.
\newblock Generative agents: Interactive simulacra of human behavior.
\newblock In {\em Proceedings of the 36th Annual ACM Symposium on User Interface Software and Technology}, pp. 1--22, 2023. \href{https://doi.org/10.1145/3586183.3606763}
{doi: {{%
10\hspace{.1pt}\discretionary{.}{%
}{.}\hspace{.4pt}1145\discretionary{/}{%
}{/}3586183\hspace{.1pt}\discretionary{.}{%
}{.}\hspace{.4pt}3606763}}}


\bibitem{chetDev}
C.~Qian, X.~Cong, C.~Yang, W.~Chen, Y.~Su, J.~Xu, Z.~Liu, and M.~Sun.
\newblock Communicative agents for software development.
\newblock {\em CoRR}, abs/2307.07924, July 2023. \href{https://doi.org/10.48550/arXiv.2307.07924}
{doi: {{%
10\hspace{.1pt}\discretionary{.}{%
}{.}\hspace{.4pt}48550\discretionary{/}{%
}{/}arXiv\hspace{.1pt}\discretionary{.}{%
}{.}\hspace{.4pt}2307\hspace{.1pt}\discretionary{.}{%
}{.}\hspace{.4pt}07924}}}


\bibitem{AgentGPT}
ReWorkd.
\newblock {AgentGPT}.
\newblock \url{https://github.com/reworkd/AgentGPT}, 2023.
\newblock Accessed on: Mar 01, 2024.

\bibitem{Salewski2023InContextIR}
L.~Salewski, S.~Alaniz, I.~Rio-Torto, E.~Schulz, and Z.~Akata.
\newblock In-context impersonation reveals large language models' strengths and biases.
\newblock In {\em Thirty-seventh Conference on Neural Information Processing Systems}, 2023. \href{https://doi.org/10.48550/arXiv.2305.14930}
{doi: {{%
10\hspace{.1pt}\discretionary{.}{%
}{.}\hspace{.4pt}48550\discretionary{/}{%
}{/}arXiv\hspace{.1pt}\discretionary{.}{%
}{.}\hspace{.4pt}2305\hspace{.1pt}\discretionary{.}{%
}{.}\hspace{.4pt}14930}}}


\bibitem{tang2023MedAgents}
X.~Tang, A.~Zou, Z.~Zhang, Y.~Zhao, X.~Zhang, A.~Cohan, and M.~Gerstein.
\newblock {MedAgents}: Large language models as collaborators for zero-shot medical reasoning.
\newblock {\em CoRR}, abs/2311.10537, Nov. 2023. \href{https://doi.org/10.48550/arXiv.2311.10537}
{doi: {{%
10\hspace{.1pt}\discretionary{.}{%
}{.}\hspace{.4pt}48550\discretionary{/}{%
}{/}arXiv\hspace{.1pt}\discretionary{.}{%
}{.}\hspace{.4pt}2311\hspace{.1pt}\discretionary{.}{%
}{.}\hspace{.4pt}10537}}}


\bibitem{Langroid}
L.~Team.
\newblock Langroid: Harness llms with multi-agent programming.
\newblock \url{https://github.com/langroid/langroid}, 2023.
\newblock Accessed on: Mar 01, 2024.

\bibitem{SuperAGI}
S.~Team.
\newblock {SuperAGI}.
\newblock \url{https://github.com/TransformerOptimus/SuperAGI}, 2023.
\newblock Accessed on: Mar 01, 2024.

\bibitem{SuperAGI-Marketplace}
S.~Team.
\newblock {SuperAGI Marketplace}.
\newblock \url{https://marketplace.superagi.com/}, 2023.
\newblock Accessed on: Mar 01, 2024.

\bibitem{wang2023survey}
L.~Wang, C.~Ma, X.~Feng, Z.~Zhang, H.~Yang, J.~Zhang, Z.~Chen, J.~Tang, X.~Chen, Y.~Lin, et~al.
\newblock A survey on large language model based autonomous agents.
\newblock {\em CoRR}, abs/2308.11432, Aug. 2023. \href{https://doi.org/10.48550/arXiv.2308.11432}
{doi: {{%
10\hspace{.1pt}\discretionary{.}{%
}{.}\hspace{.4pt}48550\discretionary{/}{%
}{/}arXiv\hspace{.1pt}\discretionary{.}{%
}{.}\hspace{.4pt}2308\hspace{.1pt}\discretionary{.}{%
}{.}\hspace{.4pt}11432}}}


\bibitem{wang2023unleashing}
Z.~Wang, S.~Mao, W.~Wu, T.~Ge, F.~Wei, and H.~Ji.
\newblock Unleashing the emergent cognitive synergy in large language models: A task-solving agent through multi-persona self-collaboration.
\newblock {\em CoRR}, abs/2307.05300, July 2023. \href{https://doi.org/10.48550/arXiv.2307.05300}
{doi: {{%
10\hspace{.1pt}\discretionary{.}{%
}{.}\hspace{.4pt}48550\discretionary{/}{%
}{/}arXiv\hspace{.1pt}\discretionary{.}{%
}{.}\hspace{.4pt}2307\hspace{.1pt}\discretionary{.}{%
}{.}\hspace{.4pt}05300}}}


\bibitem{wei2021finetuned}
J.~Wei, M.~Bosma, V.~Y. Zhao, K.~Guu, A.~W. Yu, B.~Lester, N.~Du, A.~M. Dai, and Q.~V. Le.
\newblock Finetuned language models are zero-shot learners.
\newblock In {\em The Tenth International Conference on Learning Representations}, 2022. \href{https://doi.org/10.48550/arXiv.2109.01652}
{doi: {{%
10\hspace{.1pt}\discretionary{.}{%
}{.}\hspace{.4pt}48550\discretionary{/}{%
}{/}arXiv\hspace{.1pt}\discretionary{.}{%
}{.}\hspace{.4pt}2109\hspace{.1pt}\discretionary{.}{%
}{.}\hspace{.4pt}01652}}}


\bibitem{weng2024insightlens}
L.~Weng, X.~Wang, J.~Lu, Y.~Feng, Y.~Liu, and W.~Chen.
\newblock Insightlens: Discovering and exploring insights from conversational contexts in large-language-model-powered data analysis.
\newblock {\em arXiv}, 2024. \href{https://doi.org/10.48550/ARXIV.2404.01644}
{doi: {{%
10\hspace{.1pt}\discretionary{.}{%
}{.}\hspace{.4pt}48550\discretionary{/}{%
}{/}ARXIV\hspace{.1pt}\discretionary{.}{%
}{.}\hspace{.4pt}2404\hspace{.1pt}\discretionary{.}{%
}{.}\hspace{.4pt}01644}}}


\bibitem{wong2023anchorage}
K.~K. Wong, X.~Wang, Y.~Wang, J.~He, R.~Zhang, and H.~Qu.
\newblock Anchorage: Visual analysis of satisfaction in customer service videos via anchor events.
\newblock {\em IEEE Transactions on Visualization and Computer Graphics}, 2023. \href{https://doi.org/10.48550/ARXIV.2302.06806}
{doi: {{%
10\hspace{.1pt}\discretionary{.}{%
}{.}\hspace{.4pt}48550\discretionary{/}{%
}{/}ARXIV\hspace{.1pt}\discretionary{.}{%
}{.}\hspace{.4pt}2302\hspace{.1pt}\discretionary{.}{%
}{.}\hspace{.4pt}06806}}}


\bibitem{woolley2010evidence}
A.~W. Woolley, C.~F. Chabris, A.~Pentland, N.~Hashmi, and T.~W. Malone.
\newblock Evidence for a collective intelligence factor in the performance of human groups.
\newblock {\em science}, 330(6004):686--688, Sept. 2010. \href{https://doi.org/10.1126/science.1193147}
{doi: {{%
10\hspace{.1pt}\discretionary{.}{%
}{.}\hspace{.4pt}1126\discretionary{/}{%
}{/}science\hspace{.1pt}\discretionary{.}{%
}{.}\hspace{.4pt}1193147}}}


\bibitem{wu2023autogen}
Q.~Wu, G.~Bansal, J.~Zhang, Y.~Wu, S.~Zhang, E.~Zhu, B.~Li, L.~Jiang, X.~Zhang, and C.~Wang.
\newblock {AutoGen}: Enabling next-gen llm applications via multi-agent conversation framework.
\newblock {\em CoRR}, abs/2308.08155, Aug. 2023. \href{https://doi.org/10.48550/arXiv.2308.08155}
{doi: {{%
10\hspace{.1pt}\discretionary{.}{%
}{.}\hspace{.4pt}48550\discretionary{/}{%
}{/}arXiv\hspace{.1pt}\discretionary{.}{%
}{.}\hspace{.4pt}2308\hspace{.1pt}\discretionary{.}{%
}{.}\hspace{.4pt}08155}}}


\bibitem{wu2023empiricalMath}
Y.~Wu, F.~Jia, S.~Zhang, Q.~Wu, H.~Li, E.~Zhu, Y.~Wang, Y.~T. Lee, R.~Peng, and C.~Wang.
\newblock An empirical study on challenging math problem solving with gpt-4.
\newblock {\em CoRR}, abs/2306.01337, June 2023. \href{https://doi.org/10.48550/arXiv.2306.01337}
{doi: {{%
10\hspace{.1pt}\discretionary{.}{%
}{.}\hspace{.4pt}48550\discretionary{/}{%
}{/}arXiv\hspace{.1pt}\discretionary{.}{%
}{.}\hspace{.4pt}2306\hspace{.1pt}\discretionary{.}{%
}{.}\hspace{.4pt}01337}}}


\bibitem{xagent2023}
{XAgent Team}.
\newblock {XAgent}: An autonomous agent for complex task solving.
\newblock \url{https://github.com/OpenBMB/XAgent}, 2023.
\newblock Accessed on: Mar 01, 2024.

\bibitem{xi2023rise}
Z.~Xi, W.~Chen, X.~Guo, W.~He, Y.~Ding, B.~Hong, M.~Zhang, J.~Wang, S.~Jin, E.~Zhou, et~al.
\newblock The rise and potential of large language model based agents: A survey.
\newblock {\em CoRR}, abs/2309.07864, Sept. 2023. \href{https://doi.org/10.48550/arXiv.2309.07864}
{doi: {{%
10\hspace{.1pt}\discretionary{.}{%
}{.}\hspace{.4pt}48550\discretionary{/}{%
}{/}arXiv\hspace{.1pt}\discretionary{.}{%
}{.}\hspace{.4pt}2309\hspace{.1pt}\discretionary{.}{%
}{.}\hspace{.4pt}07864}}}


\bibitem{Expertprompting}
B.~Xu, A.~Yang, J.~Lin, Q.~Wang, C.~Zhou, Y.~Zhang, and Z.~Mao.
\newblock {ExpertPrompting}: Instructing large language models to be distinguished experts.
\newblock {\em CoRR}, abs/2305.14688, May 2023. \href{https://doi.org/10.48550/arXiv.2305.14688}
{doi: {{%
10\hspace{.1pt}\discretionary{.}{%
}{.}\hspace{.4pt}48550\discretionary{/}{%
}{/}arXiv\hspace{.1pt}\discretionary{.}{%
}{.}\hspace{.4pt}2305\hspace{.1pt}\discretionary{.}{%
}{.}\hspace{.4pt}14688}}}


\bibitem{CooperativeEmbodiedAgent}
H.~Zhang, W.~Du, J.~Shan, Q.~Zhou, Y.~Du, J.~B. Tenenbaum, T.~Shu, and C.~Gan.
\newblock Building cooperative embodied agents modularly with large language models.
\newblock In {\em The Twelfth International Conference on Learning Representations}, 2024. \href{https://doi.org/10.48550/arXiv.2307.02485}
{doi: {{%
10\hspace{.1pt}\discretionary{.}{%
}{.}\hspace{.4pt}48550\discretionary{/}{%
}{/}arXiv\hspace{.1pt}\discretionary{.}{%
}{.}\hspace{.4pt}2307\hspace{.1pt}\discretionary{.}{%
}{.}\hspace{.4pt}02485}}}


\bibitem{OKRagents}
Y.~Zheng, C.~Ma, K.~Shi, and H.~Huang.
\newblock Agents meet {OKR}: an object and key results driven agent system with hierarchical self-collaboration and self-evaluation.
\newblock {\em CoRR}, abs/2311.16542, Nov. 2023. \href{https://doi.org/10.48550/arXiv.2311.16542}
{doi: {{%
10\hspace{.1pt}\discretionary{.}{%
}{.}\hspace{.4pt}48550\discretionary{/}{%
}{/}arXiv\hspace{.1pt}\discretionary{.}{%
}{.}\hspace{.4pt}2311\hspace{.1pt}\discretionary{.}{%
}{.}\hspace{.4pt}16542}}}


\bibitem{zhuge2023mindstorms}
M.~Zhuge, H.~Liu, F.~Faccio, D.~R. Ashley, R.~Csord{\'a}s, A.~Gopalakrishnan, A.~Hamdi, H.~A. A.~K. Hammoud, V.~Herrmann, K.~Irie, et~al.
\newblock Mindstorms in natural language-based societies of mind.
\newblock {\em CoRR}, abs/2305.17066, May 2023. \href{https://doi.org/10.48550/arXiv.2305.17066}
{doi: {{%
10\hspace{.1pt}\discretionary{.}{%
}{.}\hspace{.4pt}48550\discretionary{/}{%
}{/}arXiv\hspace{.1pt}\discretionary{.}{%
}{.}\hspace{.4pt}2305\hspace{.1pt}\discretionary{.}{%
}{.}\hspace{.4pt}17066}}}


\end{thebibliography}

\appendix 

\end{document}